\documentclass[twocolumn]{aastex63}

\usepackage{amsmath}
\usepackage{longtable}
\usepackage{graphicx}

\definecolor{amethyst}{rgb}{0.6, 0.4, 0.8}
\definecolor{byzantine}{rgb}{0.74, 0.2, 0.64}

\defcitealias{bleem20}{B20}

\begin{document}

\title{Asteroid Measurements at Millimeter Wavelengths with the South Pole Telescope}

\author[0000-0002-5397-9035]{P.~M.~Chichura} \affiliation{Department of Physics, University of Chicago, 5640 South Ellis Avenue, Chicago, IL, 60637, USA} \affiliation{Kavli Institute for Cosmological Physics, University of Chicago, 5640 South Ellis Avenue, Chicago, IL, 60637, USA}
\author[0000-0002-7145-1824]{A.~Foster} \affiliation{Department of Physics, Case Western Reserve University, Cleveland, OH, 44106, USA}
\author{C.~Patel} \affiliation{Department of Physics, University of Illinois Urbana-Champaign, 1110 West Green Street, Urbana, IL, 61801, USA}
\author{N.~Ossa-Jaen} \affiliation{Department of Physics, University of Redlands, 1267 East Colton Ave, Redlands, CA, 92374, USA}
\author{P.~A.~R.~Ade} \affiliation{School of Physics and Astronomy, Cardiff University, Cardiff CF24 3YB, United Kingdom}
\author{Z.~Ahmed} \affiliation{Kavli Institute for Particle Astrophysics and Cosmology, Stanford University, 452 Lomita Mall, Stanford, CA, 94305, USA} \affiliation{SLAC National Accelerator Laboratory, 2575 Sand Hill Road, Menlo Park, CA, 94025, USA}
\author{A.~J.~Anderson} \affiliation{Fermi National Accelerator Laboratory, MS209, P.O. Box 500, Batavia, IL, 60510, USA} \affiliation{Kavli Institute for Cosmological Physics, University of Chicago, 5640 South Ellis Avenue, Chicago, IL, 60637, USA}
\author{M.~Archipley} \affiliation{Department of Astronomy, University of Illinois at Urbana-Champaign, 1002 West Green Street, Urbana, IL, 61801, USA} \affiliation{Center for AstroPhysical Surveys, National Center for Supercomputing Applications, Urbana, IL, 61801, USA}
\author{J.~E.~Austermann} \affiliation{NIST Quantum Devices Group, 325 Broadway Mailcode 817.03, Boulder, CO, USA 80305} \affiliation{Department of Physics, University of Colorado, Boulder, CO, USA 80309}
\author{J.~S.~Avva} \affiliation{Department of Physics, University of California, Berkeley, CA, 94720, USA}
% ...

\shortauthors{Chichura et al.}
\shorttitle{SPT Asteroids}

\begin{abstract}
We present the first measurements of asteroids in millimeter wavelength (mm) data from the South Pole Telescope (SPT), which is used primarily to study the cosmic microwave background (CMB).
We analyze maps of two $\sim270$ deg$^2$ sky regions near the ecliptic plane, each observed with the SPTpol camera $\sim100$ times over one month.
We subtract the mean of all maps of a given field, removing static sky signal, and then average the mean-subtracted maps at known asteroid locations.
We detect three asteroids---(324) Bamberga, (13) Egeria, and (22) Kalliope---with signal-to-noise ratios (S/N) of 11.2, 10.4, and 6.1, respectively, at 2.0~mm (150~GHz); we also detect (324) Bamberga with S/N of 4.1 at 3.2~mm (95~GHz).
We place constraints on these asteroids’ effective emissivities, brightness temperatures, and light curve modulation amplitude.
Our flux density measurements of (324) Bamberga and (13) Egeria roughly agree with predictions, while our measurements of (22) Kalliope suggest lower flux, corresponding to effective emissivities of $0.64 \pm 0.11$ at 2.0~mm and $<0.47$ at 3.2~mm.
We predict the asteroids detectable in other SPT datasets and find good agreement with detections of (772) Tanete and (1093) Freda in recent data from the SPT-3G camera, which has $\sim10 \times$ the mapping speed of SPTpol.
This work is the first focused analysis of asteroids in data from CMB surveys, and it demonstrates we can repurpose historic and future datasets for asteroid studies.
Future SPT measurements can help constrain the distribution of surface properties over a larger asteroid population.
\end{abstract}

 
\section{Introduction}
\label{sec:intro}

Astronomers can learn about the evolution of our solar system and its planets by studying the physical properties of asteroids \citep{michel15}.
Typically, astronomers observe asteroids passively at optical and thermal infrared (IR) wavelengths; in these wavelength ranges, the asteroid flux densities are dominated by reflected solar light and thermal emission, respectively.
Astronomers also study asteroids actively with radar observations, which detect echo signals to determine physical shape and properties.
While astronomers do study asteroids at sub-millimeter (submm), millimeter (mm), and centimeter (cm) wavelengths, such studies are less frequent, despite the feasibility of such measurements having been demonstrated as early as the 1970s \citep{briggs73}.
These microwave observations provide information that optical, IR, and radar observations cannot.
For instance, early observers at microwave wavelengths correctly understood that emission from wavelengths longer than IR originated from multiple wavelengths into the ``regolith,'' i.e. the unconsolidated rocky surface, of the asteroid depending on regolith composition \citep{ulich76,conklin77,johnston82}.
That is, the regolith becomes more transparent at longer wavelengths, so early observers found they could measure thermal radiation emitted from deeper under the asteroid's surface.

Studies at IR wavelengths suggested that most large asteroids had surface emissivities near unity, yet early observations from submm to cm wavelengths measured flux densities much lower than models predicted.
These early observers interpreted the lower flux densities as the result of a wavelength-dependent drop in emissivity as large as 25\% \citep{johnston82,webster88}.
At the time, astronomers interpreted this lower emissivity as an effective emissivity resulting from scattering by grains within the regolith, which would make it harder for photons to escape to the surface \citep{redman92}.
Astronomers used this interpretation to place constraints on the composition and properties of asteroids' surfaces, and they continued interpreting asteroids in this way for decades, including in some recent studies at these wavelengths that observe emissivity drops as great as 40\% \citep{muller07,moullet10}.

However, there is mounting evidence that this interpretation is incorrect.
The European Space Agency's \textit{Rosetta} is the first Solar System spacecraft mission that includes instrumentation able to measure thermal fluxes at IR, submm, and mm wavelengths.
\textit{Rosetta} made close approaches to two asteroids, one of which was the large asteroid (21) Lutetia\footnote{The naming convention of asteroids consists of an object's International Astronomical Union designation number in parentheses followed by its name (if any).} \citep{gulkis12}.
During the flyby of (21) Lutetia, \textit{Rosetta} also recorded a decrease in flux at mm wavelengths, but more complex modeling suggested that this was due to a large temperature gradient in the outer regolith as opposed to a wavelength-dependent emissivity \citep{keihm12}.
Later, \citet{keihm13} applied their modeling to thermal fluxes of other large asteroids to suggest an altogether new interpretation of the observed decrease in flux at longer wavelengths.
They found that the decrease in flux could be explained by emissivities near unity at all wavelengths combined with a significant temperature gradient over depth, with temperatures as much as 50-80~K lower several mm below the asteroid's surface.

The new interpretation suggested by \citet{keihm13} represents a paradigm shift that would fundamentally alter the way astronomers examine asteroid regolith composition.
In order to expand on this work, astronomers need more high-sensitivity measurements of asteroids at submm to cm wavelengths, where observations exist for only a handful of large asteroids.
Recently, the Atacama Large Millimeter/submillimeter Array (ALMA) has carried out dedicated studies of asteroids and other small Solar System bodies, including asteroids (1) Ceres \citep{li2020}, (3) Juno \citep{alma15}, and (16) Psyche \citep{dekleer21}, as well as Centaurs and trans-Neptunian objects \citep{lellouch17}.
Measurements like these at submm to cm wavelengths serve an important role in asteroid studies because they lie on the ill-understood boundary between two observable regimes---the highly emissive radiation in IR and the supposedly less emissive radiation in cm---and ultimately can improve modeling of surface properties, including thermal inertia and regolith roughness.

Instruments like ALMA can track celestial targets for a short time with high sensitivity, but these instruments are generally facilities which require proposals for use.
These facilities receive many observation requests, so studying asteroids comes at a high opportunity cost.
However, astronomers can incur no opportunity cost if they repurpose data from other types of observations that happen to include asteroids.
Sky surveys at mm wavelengths fill this niche and are made frequently using telescopes designed to study the cosmic microwave background (CMB).

In this paper, we show that we can use data from the South Pole Telescope (SPT) to detect asteroids at high signal-to-noise ratios (S/N) at mm wavelengths when we average maps of the sky centered on known asteroid locations.
By showing this, we demonstrate that historic and future data from CMB experiments can be repurposed for observing asteroids.
In Section \ref{sec:inst}, we describe the SPT and the cameras from which the data in this paper are taken, the specific observations used in this work, and the data processing used in making the single-observation maps used in the asteroid search.
In Section \ref{sec:query}, we explain the asteroid selection criteria in historic data.
In Section \ref{sec:detect}, we describe how we search for the selected asteroids in our observations.
In Section \ref{sec:results}, we show the detection of three large asteroids---(324) Bamberga, (13) Egeria, and (22) Kalliope---with the SPTpol camera at 2~mm (150~GHz), as well as (324) Bamberga at 3.2~mm (95~GHz).
In Section \ref{sec:discuss}, we discuss these results.
In Section \ref{sec:pred}, we suggest prospects for continuing this analysis on other data sets.
We conclude in Section \ref{sec:conc}.

Although the \textit{Planck} collaboration has previously published detections of asteroids in their analysis connecting dust observations to asteroid families \citep{cremonese02, planck13-14} and the Atacama Cosmology Telescope (ACT) collaboration masked asteroids in their search for Planet 9 \citep{naess2021b}, this work represents the first focused analysis of asteroid flux in data from an experiment designed to measure the CMB.
With continued analysis, historic and future data measuring the CMB can provide more observations of asteroids at submm and mm wavelengths.
In particular, scientists can make these measurements using a wealth of data provided by current experiments, such as those on the SPT and the ACT, as well as upcoming experiments like the Simons Observatory (SO) and CMB-S4 \citep{kosowsky03,ade19,abazajian16}.

\section{Instruments, Observations, Data Processing}
\label{sec:inst}
The primary results in this work use observations from the SPTpol camera on the SPT, 
with some proof-of-concept results from the currently installed SPT-3G camera.
We provide a brief description of the telescope and both cameras, the method of data collection---particularly
as it pertains to the sensitivity to moving objects---and the standard data analysis through the mapmaking
step. We describe the post-map processing specific to asteroid detection and
characterization in Section~\ref{sec:detect}.

\subsection{Telescope and Cameras}
\label{sec:telcam}
The SPT is a mm/submm telescope with a 10-meter primary
mirror, installed at the National Science Foundation Amundsen-Scott South Pole research station.
The telescope is physically located approximately 1~km from the geographical South Pole.
Since its construction in 2006-2007, the SPT has been used almost exclusively to make deep
maps of thousands of square degrees of the Southern sky, with the primary science goal of 
characterizing the primary and secondary CMB anisotropies in intensity and polarization. 
For more details on the telescope, see \citet{carlstrom11} and \citet{padin08}.

SPTpol was the second camera installed on the telescope, replacing the original SPT-SZ camera
in 2012. SPTpol consisted of 1536 feedhorn-coupled, polarization-sensitive superconducting detectors, 
1176 sensitive to radiation in a band centered near 2.0~mm (150~GHz) and 360 sensitive to radiation 
in a band centered near 3.2~mm (95~GHz). Although we refer to these bands as ``2.0~mm'' and ``3.2~mm,''
the band centers are closer to 2.01~mm (149.3~GHz) and 3.11~mm (96.2~GHz), respectively, for a
Rayleigh-Jeans spectrum, like that expected from asteroids.  These effective band centers may shift slightly
if we consider the drop in effective emissivity described in Section \ref{sec:intro}, but this shift would only minimally alter
measured fluxes and effective emissivities.
We can approximate the main
lobes of the SPTpol beams or point-spread functions in the two bands by Gaussians
with FWHM equal to roughly $1$\farcm$2$ at 2.0~mm and $1$\farcm$6$ at 3.2~mm.
More details on SPTpol can be found in \citet{austermann12} and \citet{bleem12a}.

The SPT-3G camera replaced SPTpol on the telescope in 2017. SPT-3G consists of 
$\sim$16,000 superconducting detectors configured to observe in three bands, centered at
roughly 1.4~mm (220~GHz), 2.0~mm (150~GHz), and 3.2~mm (95~GHz). Each camera pixel
is coupled to two (orthogonally polarized) detectors in each of the three bands.
The beam FWHM for SPT-3G is similar to that in SPTpol for 
the two common bands and is roughly $1$\farcm$05$ at 1.4~mm. For more details on SPT-3G, see
\citet{anderson18} and \citet{sobrin22}.

Both SPTpol and SPT-3G contain polarization-sensitive detectors.
Observers might expect to measure polarization of microwave emissions but only at polarization levels of a few tens of percents \citep{lagerros96}.
Indeed, recent observations at mm wavelengths have found polarization levels even lower than theory might suggest \citep{dekleer21}.
Therefore, we expect that any measurements of polarized light from the asteroids considered in this paper would be approximately an order of magnitude weaker than measurements of total intensity.
Given the significance at which we ultimately detect unpolarized emission of asteroids considered in this paper, we only use the total intensity information from SPTpol and SPT-3G in this work.

\subsection{Observations}
\label{sec:obs}
The primary results in this work come from observations with the SPTpol camera of two sky regions: ``\textsc{ra13hdec-25}'' centered at roughly right ascension (R.A.)~$13^h$, declination (decl.)~$-25^\circ$, and ``\textsc{ra23hdec-25}'' centered at roughly R.A.~$23^h$, decl.~$-25^\circ$. Each field is $2^h$ wide in R.A.~and $10^\circ$ tall in decl.,~covering
roughly 270 square degrees each.
We note that these are different from the primary SPTpol science field, a 500-square-degree patch centered at R.A.~$0^h$, decl.~$-57.5^\circ$ \citep{henning18}.

From roughly December through March, the primary science field was partially contaminated by the Sun due to telescope sidelobes, so SPTpol was used to conduct 
a supplementary survey of other fields with relatively low Galactic foreground emission. This supplementary survey
is called the SPTpol Extended Cluster Survey (ECS), the details of which can be found 
in \citet[][hereafter B20]{bleem20}. The ECS covers nearly 2800 square degrees; we concentrate on the 
two fields mentioned above because of their proximity to the Ecliptic plane, as explained in Section
\ref{sec:query}. 

Each of these two fields was observed for roughly one month in either 2015 or 2016, on a 
roughly 2.5-hour cadence, with a $\sim$4-hour pause every $\sim$24 hours
for cycling the helium adsorption refrigerator
that cools the detectors. 
The \textsc{ra23hdec-25} field was observed from 2015 October 29 to 
2015 November 29.
The mean position of the sun during this time was at R.A.~$226.6^\circ$ and decl.~$-17.2^\circ$.
The \textsc{ra13hdec-25} field was observed for one day on 2016 February 13 then 
from 2016 February 23 through 2016 March 22. 
The mean position of the sun during this time was at R.A.~$315.9^\circ$ and decl.~$-4.2^\circ$.
\citetalias{bleem20} estimates the final noise level of the two fields to be roughly 30 $\mu$K-arcmin at
2.0~mm and 50-60 $\mu$K-arcmin at 3.2~mm, corresponding to 1$\sigma$ point-source sensitivities
of roughly 2 and 3~mJy at 2.0~mm and 3.2~mm, respectively.

In Section~\ref{sec:pred}, we perform a rough validation of predictions for asteroid yield in other
surveys using 2020 data from the main SPT-3G science field, a 1500-square-degree field centered
at R.A.~$0^h$, decl.~$-56^\circ$ (a superset of the SPTpol main science field). In 2020 this field was
observed from March through November, with an effective cadence of 1-2 days (the full field is split
into four subfields, two of which are observed during a given refrigerator cycle---see 
\citealt{guns21}
for details).

\subsection{Data Processing}
\label{sec:dataproc}

\subsubsection{TOD Filtering and Mapmaking}
\label{sec:maps}
The maps used in this work were originally created for use in the cluster-finding analysis of 
\citetalias{bleem20}. For details of the data processing used to make these maps, we refer
the reader to that work; we summarize the basic steps here. 
For each observation, the time-ordered data (TOD) were subject to quality cuts and several 
filtering steps, including bandpass filtering to suppress low- and high-frequency noise and 
common-mode subtraction to suppress atmospheric contamination.
The sky location to which each detector was pointed at each time sample was then calculated 
and binned into $0$\farcm$25$ pixels in the Sanson-Flamsteed projection. 
Finally, all time samples from all detectors in a given
observing band pointing to a given pixel were averaged with inverse-noise-weighting
to produce the map.

Because of
the finite resolution of the telescope and the filtering applied to the TOD, the resulting maps are biased
representations of the true sky signal. We can represent both the telescope resolution effect
and the effect of the TOD filtering as multiplications in two-dimensional Fourier
space, and we refer to these as the beam $B(\mathbf{l})$ and the filter transfer function $F(\mathbf{l})$, 
respectively, where $\mathbf{l} \in \{l_x, l_y\}$ is the wavenumber equivalent of the Fourier coordinate system $\{u,v\}$
(i.e., $l_x = 2 \pi u$, $l_y = 2 \pi v$). The most important TOD filtering steps are: 1) a scan-direction high-pass
filter that, combined with the azimuth-raster scan strategy and the polar telescope location, results in a
map-space $x$-direction high-pass with a cutoff of $l_x = 300$; and 2) a common-mode subtraction across
the array at each time step that acts in map space as an isotropic high-pass with a cutoff of roughly 
$\ell \equiv |\mathbf{l}| = 300$.

\subsubsection{Calibration}
\label{sec:cal}
Relative gains across the detector array and between observations are measured using regular observations of the Galactic HII region RCW38 and regular observations of an internal thermal calibration source.
As in \citetalias{bleem20}, the absolute calibration of the map was derived by comparing the full-season coadded maps with the \textit{Planck} map of the same field.
The SPT-ECS fields were taken at significantly higher levels of atmospheric loading compared to other SPTpol 
survey data, and the resulting larger change in detector loading with elevation necessitated a further calibration
step beyond a constant normalization factor for 3.2~mm data. 
Although noise in SPTpol data does not in general depend strongly on airmass, the 3.2~mm data required this additional calibration step as the calibration was empirically found to vary 
significantly with elevation within a field (which is equivalent to decl.~for observations from the South Pole). 
This trend is fit well as a linear function of decl.,~and \citetalias{bleem20} used the \textit{Planck} data to fit for and correct 
this variation across the fields.

Because we aim in this work to measure asteroid flux density on an observation-to-observation basis, 
we also compare individual observations with \textit{Planck} data.
We find that the single-observation calibration varies significantly less than the noise on the measurement of any asteroid
in a single observation, and we ignore this as a source of uncertainty in subsequent analysis.
We also repeat the B20 comparison of the full-depth SPTpol maps with \textit{Planck} data and make small 
($\sim$5\%-level) corrections to the B20 absolute calibration.

The comparison to \textit{Planck} yields a calibration at angular scales where the CMB temperature anisotropy
is strongest (roughly degree scales), and transferring this to a flux density scale requires accurate knowledge of 
the beam and filtering.
We verify our flux density calibration by comparing flux measurements of the source J2258-2758 at 3.2~mm in \textsc{ra23hdec-25} with the value reported in the ALMA Calibrator Source Catalogue.\footnote{https://almascience.nrao.edu/sc/}
This source is the only source recorded in the catalogue during the observation periods of our fields.
We found that ALMA's measurement was within $\sim1$ standard deviation of our measurement from 5 observations near ALMA's observation date. 
Because we are primarily checking for systematic failures in the beam and filter transfer function estimation, 
we take the success of the verification at 3.2~mm and in one field to indicate that the flux density calibration
chain is likely robust at 2.0~mm and in the other field.

Overall, these calibration steps carry a few systematic uncertainties.
Uncertainty in 3.2~mm data from the elevation-dependent recalibration done by \citetalias{bleem20} is around 5.9\% (5.5\%) for \textsc{ra13hdec-25} (\textsc{ra23hdec-25}).
Uncertainty from calibration with \textit{Planck} data is around 1.4\% (2.4\%) for \textsc{ra13hdec-25} (\textsc{ra23hdec-25}) in 2.0~mm data and 2.3\% (2.0\%) for \textsc{ra13hdec-25} (\textsc{ra23hdec-25}) in 3.2~mm data.
There is further uncertainty related to the SPTpol beam shape used in filtering maps and converting map units to flux units; this is at most on the scale of a few percent.
Added in quadrature, these systematic uncertainties are roughly 6\% for the 3.2~mm data and 3\% for the 2.0~mm data, which are subdominant to noise fluctuations in observations of the asteroids reported in Section \ref{sec:results}.

\section{Selecting Asteroids to Examine}
\label{sec:query}

\begin{figure*}[ht!]
	\centering
	\includegraphics[clip,trim=0 {0.092\linewidth} 0 {0.092\linewidth},width=0.8\linewidth]{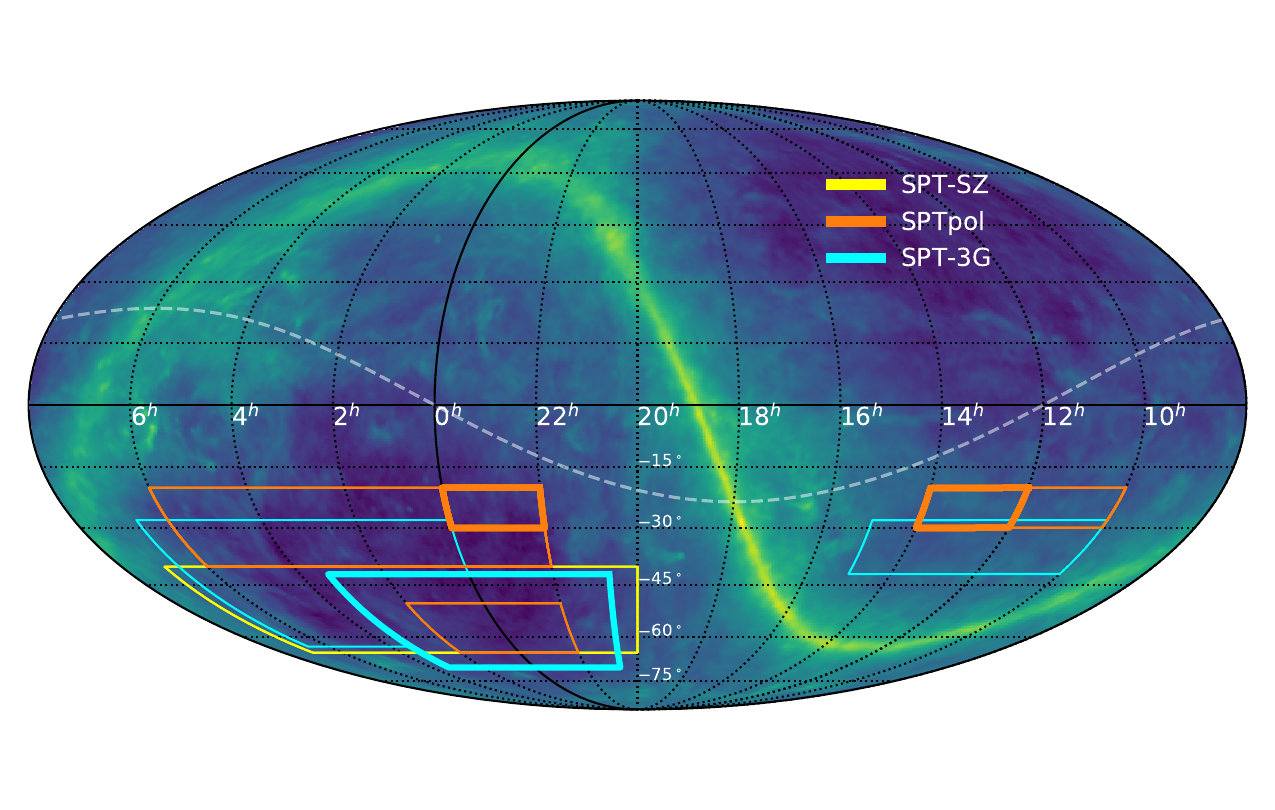}
	\caption{SPT-SZ, SPTpol, and SPT-3G observation fields plotted on a Mollweide projection of the sky using the equatorial coordinate system.
	The yellow, orange, and cyan boxes denote the sky regions observed using the SPT, and thicker boxes denote regions analyzed in this paper.
	The blue-green-yellow color scale represents galactic dust emission at 545 GHz from the Planck Public Data Release 2, with yellow indicating higher emission \citep{planck15-10}.
	The dashed gray line represents the ecliptic plane.
	}
	\label{fig:footprints}
\end{figure*}

Figure \ref{fig:footprints} shows a map of the Galactic dust emission from the \emph{Planck} satellite, with the locations of the SPTpol and SPT-3G observation fields and the ecliptic plane superimposed.
Galactic emission can obscure measurements of the CMB, so CMB survey designers typically choose observation fields that avoid this emission.
The ecliptic plane marks the apparent path of the Sun through the sky over the course of a year, so near it we should find  objects in our solar system with low orbital inclination, like most main belt asteroids (MBA).
Thus, we would expect to find more observable asteroids in the fields closest to the ecliptic, and we focus our initial search on those fields: SPTpol ECS fields \textsc{ra13hdec-25} and \textsc{ra23hdec-25} (detailed in Section \ref{sec:obs}).

To identify observable asteroids, we predict the asteroids' flux densities using standard thermal models.
Three such models are commonly employed: the ``Standard Thermal Model'' (STM) developed by \citet{lebofsky86}, the ``Fast Rotating Model'' (FRM) developed by \citet{lebofsky89}, and the ``Near Earth Asteroid Thermal Model'' (NEATM) developed by \citet{harris98}.
For a review of all three models, we refer the reader to \citet{delbo02}.
All three models consider thermal equilibrium between incoming solar radiation and outgoing emitted radiation on the surface of an asteroid.
STM considers the case in which an asteroid is not rotating, reaches a maximum temperature at the subsolar point, and radiates only from the day side.
FRM considers the case in which an asteroid is rotating quickly and reaches a maximum temperature along its hemisphere.
NEATM considers the same case as STM, except it integrates emission from the asteroid's visible surface as opposed to using empirical scaling with solar phase angle.
The models generally take the following parameters: solar distance, Earth distance, solar phase angle, bond albedo, emissivity, and ``beaming parameter.''
The beaming parameter was originally included to account for an effect called ``infrared beaming'' in which thermal emission is greater from a rough surface when viewed at small phase angles, but it is more often treated as a way to empirically scale the model.
For STM, the beaming parameter can vary from asteroid to asteroid but is assumed constant over wavelength, and the value of 0.756 was determined empirically by \citet{lebofsky86} from thermal IR measurements of (1) Ceres and (2) Pallas.
For FRM, the beaming parameter is fixed at $\pi$.
For NEATM, the beaming parameter is a free parameter which varies to fit the data.
STM with 0 solar phase angle is the equivalent of a non-rotating spherical blackbody and predicts the maximum possible flux density for given observation geometry and physical parameters, while FRM is considered the minimum possible.
NEATM has become the standard model for thermal asteroid analysis following its wide use by the NEOWISE collaboration \citep{mainzer11}.

To identify asteroids present in those fields, we compiled a list of potentially visible objects.
First, we queried the JPL Small-Body Database (SBD) Search Engine \citep{giorgini20} to generate a list of all small bodies with reported values of effective spherical diameter $D$, perihelion distance $q$, and minimum orbit intersection distance (MOID).
Next, we estimated the maximum possible flux density of each of those objects with the STM.
We considered maximum possible flux by assuming values of 1 for emissivity and 0 for albedo, and using $q$ as solar distance and MOID as Earth distance for optimal viewing geometry.
We adopted the standard empirical assumption of 0.756 for the model's ``beaming parameter.''
Finally, we eliminated all asteroids for which the maximum possible flux fell below 0.5~mJy, which would correspond to roughly a $2\sigma$ fluctuation at the projected depth of the main SPT-3G field after 5 years of observing.
However, we maintained all comets and near Earth objects as potentially interesting objects regardless of maximum possible flux.
This resulted in a list of 5,885 objects of interest.

After compiling a list of viable observation candidates, we identified those that were within our observation fields during our observation periods.
We did so using \texttt{astroquery}, a set of Python tools developed by \citet{ginsburg19} to request data from online astronomical databases and web services.
First, we queried JPL HORIZONS Web-Interface \citep{giorgini96} to generate ephemerides for each asteroid on each day we observe in each field.
Next, we filtered the list of objects by removing all those that were outside of the fields during the entire observation periods.
Then, we queried the JPL SBD for each object's diameter and albedo, and we used these values and their ephemerides in the STM to estimate the objects' fluxes.
Finally, we estimated the expected noise levels by scaling the fully integrated noise levels reported in \citetalias{bleem20} depending on the number of observations in which each asteroid was present.
We computed a prediction of S/N by dividing each asteroid's average flux density by its estimated noise level.
Although we attempted to detect all 136 objects that passed through our fields, we only present results here for those 3 asteroids with a predicted S/N at 2.0~mm greater than 5.

We note that these predicted S/N values are likely upper limits due to multiple assumptions that affect the asteroids' temperatures.
For one, the STM models asteroid \textit{surface} temperatures, and we expect that mm emission originates from cooler, subsurface regions \citep{keihm13}.
Likewise, the STM models the limit of a non-rotating asteroid, in which case the asteroid reaches maximum possible temperature, so we expect that a realistic, rotating asteroid would be cooler.
Predicted S/N values will also differ from measured values depending on the asteroids' rotations relative to the observer.
That is, an asteroid whose hot subsolar point is rotating away from the observer is in thermal ``morning'' and will appear less bright than an asteroid whose hot subsolar point is rotating toward the observer in thermal ``afternoon.''

\section{Methods for Detecting Asteroids and Constraining Their Properties}
\label{sec:detect}

Using the methods of Section \ref{sec:query}, we compile a list of asteroids that are known to pass through our fields during our observation periods.
In this section, we describe the methods we used to measure emission from the selected asteroids in SPT data.

First, we calculate the noise in each individual-observation map, which were previously constructed from $\sim$2.5-hr long observations of each field.
We also compare the apparent positions of bright extragalactic sources in each individual-observation map with the known positions of those sources in the AT20G catalog to double-check astrometry \citep{murphy10}.
We then ``coadd'' the individual-observation maps of each field, excluding any observations that were outliers in the distribution of map noise or astrometry.
For details on the coadding process, we refer readers to \citetalias{bleem20} and \citet{everett20}.
A coadded map measures the sky’s average signal; each pixel’s value in the coadded map is the average of that pixel from the input maps weighted by inverse variance.
Because the coadded maps are averages, we retain power dominated by static sources, but we average out power from variable sources and moving sources, including asteroids in particular.
Next, we subtract the coadded map from each individual-observation map to create ``differenced maps.''
Because we subtracted off the power from static sources, these differenced maps should contain only noise and flux from transient and variable sources.
When subtracting the coadded map which includes transient and variable sources, we do introduce a bias by removing some of the sources' power, but this bias is at the percent level since no asteroid observation contributes more than $\sim1.5\%$ to the coadded map.
This bias would be much larger for asteroids moving at angular speeds much slower than the size of one SPTpol beam between observations, or roughly 29$\arcsec$~$\mathrm{hr}^{-1}$.
In this work we analyze asteroids moving quickly enough that this bias is not concerning, with (324) Bamberga moving the slowest at an average angular speed of $\sim$20$\arcsec$~$\mathrm{hr}^{-1}$. 
Because our goal is to detect asteroids, the differenced maps are the primary form of data that we analyze in this work.

To enhance the sensitivity of the asteroid search, we choose to look at the locations of known asteroids, which requires precisely knowing those locations.
Some asteroids can move across the sky at angular speeds such that the change in their position over an hour is comparable to the SPT beam size.
Since each observation lasts roughly 2 hours, we must more precisely define what time we scan over any asteroid.
We maintain some precision by considering the SPT's scanning strategy, which involves scanning back and forth in azimuth before stepping in elevation, which at the South Pole is equivalent to stepping in decl.
If we know an asteroid's decl.~around the time of observation, we can interpolate a more precise time at which we scan over the asteroid.
First, we queried JPL HORIZONS Web-Interface using \texttt{astroquery} to generate ephemerides for all asteroids at the time halfway through each observation, an initial guess.
Then, we estimated the time at which the telescope would scan over the asteroid and re-queried JPL HORIZONS Web-Interface to obtain a more precise location.
Given typical MBA motions, we assume our positional errors to be much less than the SPT beam size.

Using these more precise asteroid locations, we cut out small regions of each differenced map centered on the asteroid location.
We conduct multiple analyses on these cutouts.
To report mean flux measurements, we coadd the cutouts and filter the coadded cutout with a ``matched filter'' that maximizes the S/N for point sources.
For details on matched filtering SPT data, we refer the reader to \citet{everett20}.
The resulting measurements are in units of $T_\mathrm{CMB}$; i.e. map values are expressed as equivalent fluctuations from the mean CMB blackbody temperature of 2.726~K.
We convert the value of the center pixel to report flux density in units of mJy.
We calculate uncertainties and S/N by computing the root mean square (rms) noise in areas of the coadded cutout between $1$\farcm$5$ and $15$\arcmin$ $ away from the asteroid.
We report our mean flux measurements in Section \ref{sec:results}.

We can calculate other useful information from mean flux measurements.
To do so, we use the NEATM to remain consistent with standard reporting of thermal emission measurements.
NEOWISE reported diameters, albedos, and beaming parameters for thousands of asteroids, and we use those values to predict our asteroids' expected fluxes more reliably \citep{mainzer19}.
Once we compute the NEATM flux predictions, we can calculate effective emissivities in each band by dividing the measured flux by the model flux.
Likewise, we can use the NEATM to solve for the sub-solar temperature which would produce the fluxes we measure, and we scale these temperatures by solar distance $r_\odot$ according to the NEATM's assumed dependence of $r_\odot^{-1/2}$.
Not only can this brightness temperature be viewed as another way to report mean flux density, but calculated brightness temperature and effective emissivity can also be viewed as probes of the long-wavelength emission drop described in Section \ref{sec:intro}.
Furthermore, because we make flux measurements at multiple wavelengths, we can calculate the spectral index $\alpha$ between the two wavelengths.
We define $\alpha$ in terms of measured flux $S_\nu$ at frequency $\nu$ such that
\begin{equation}
	S_\nu \sim \nu^\alpha
\end{equation}
so that it is easy to compare to expected thermal emission with $\alpha=2$.
We report our predicted fluxes, calculated effective emissivities, calculated brightness temperatures, and calculated spectral indices in Section \ref{sec:results}.

SPTpol does not have high enough sensitivity to detect most asteroids with high significance in individual observations, but we can still consider the flux of the center pixel as a function of time, i.e. the target's light curve.
We create the light curves by matched filtering each non-coadded differenced map and plotting flux versus time of observation.
In this paper, we create the light curves only for 2.0~mm data because our observations at this wavelength have higher S/N and do not require additional elevation-dependence calibration like our 3.2~mm data.
Light curves are important because observed flux density changes depending on viewing geometry.
We test whether we detect the expected change in flux density by calculating the difference in $\chi^2$ between models that consider only constant flux from the source versus constant flux plus variation as predicted by the NEATM.

Asteroids in general are not spherical, so as they rotate while traveling through space, we expect to observe a periodic modulation in their light curves.
If we can detect modulation of this type in the light curves, we can infer information about an asteroid's shape, rotational period, and other properties.
First, we scale the light curves by a correction factor to account for flux changes due to viewing geometry.
This is done by calculating the mean flux predicted by the NEATM and scaling the light curve to that value.
For a sense of what that scaling might look like, consider the STM in the long-wavelength Rayleigh-Jeans limit, in which case flux density $F$ varies like
\begin{equation}
	F \sim r_{\odot}^{-1/2} r_{\oplus}^{-2} 10^{-0.004 \alpha}
\end{equation}
where $r_{\odot}$ is solar distance to the asteroid, $r_{\oplus}$ is Earth distance to the asteroid, and $\alpha$ is solar phase angle measured in degrees, an empirical fit.
We compute scaling factors to mean values of $r_{\odot}$, $r_{\oplus}$, and $\alpha$ based on NEATM predictions, which is comparable at the percent level to using the above functional form.
Next, we compute Lomb-Scargle periodograms to try detecting statistically significant periods.
For details on the periodogram, we refer readers to \citet{vanderplas18}.
Finally, if the asteroid has a known period, we can also ``fold'' the light curve by plotting flux versus observation time modulo rotation period.
When we fold the light curve in this way, we plot flux as a function of rotational phase, and we fit a sinusoidal function to place limits on modulation amplitudes at mm wavelengths.
To first order, asteroids are ellipsoids, so we would expect the most observable modulation to be sinusoidal with a period half that of the asteroid's known rotational period.

One caveat to our light curve analysis is that each asteroid is scanned over multiple times in each $\sim$2.5-hour individual observation. This introduces two effects: an inexactness for time observed and an averaging of flux from the asteroid during that time. If the rotation period of the observed asteroid is short compared to the timescale over which the asteroid is observed, these effects will both reduce sensitivity to the brightness modulation induced by rotation. For the SPTpol observations under study, in which the telescope is scanned back and forth in azimuth then stepped in elevation, the relevant observation timescale is the time during which a particular sky elevation is within the field of view of the camera. The elevation extent of the SPTpol camera is roughly 1 degree; thus, in a $\sim$2.5-hour observation of a 10-degree field, a particular elevation will be visible for approximately 15 minutes. This is much less than the multiple-hour rotational periods of our asteroids of interest, and we conclude that our sensitivity to rotational effects is not compromised by these effects.\footnote{We note that we could restore nearly full sensitivity to changing asteroid brightness by analyzing single telescope scans individually, as was done in \citet{guns21}, but the scaling arguments above imply this would not improve our sensitivity to asteroid brightness changes in any material way.}

\section{Results}
\label{sec:results}
	
\begin{figure*}[]
	\centering
	\includegraphics[width=0.8\linewidth]{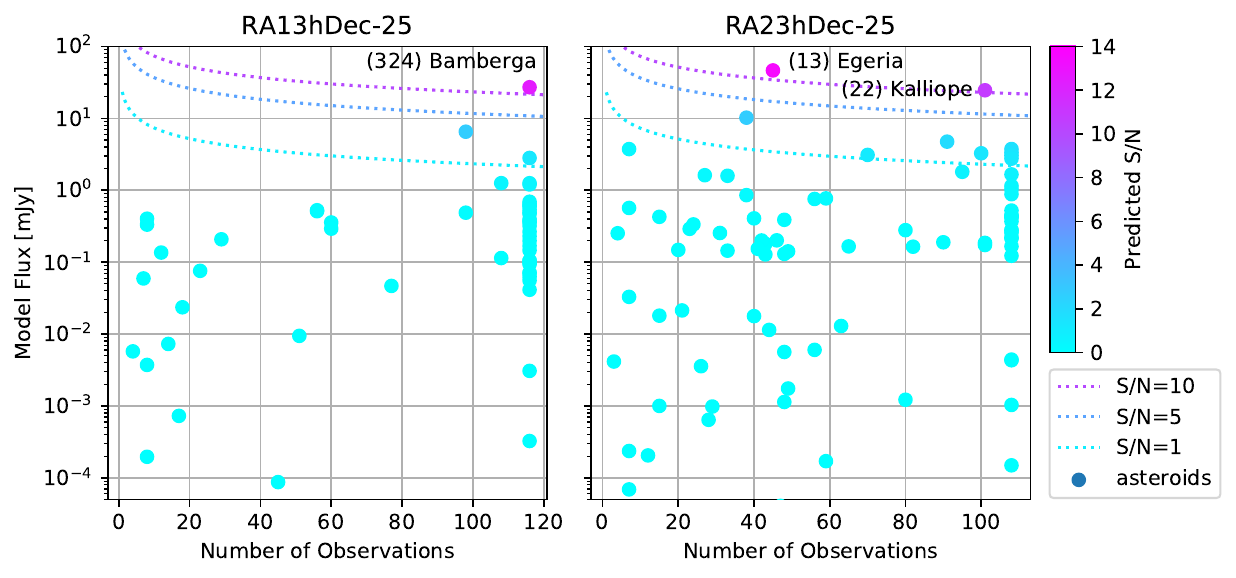}
	\caption{Predicted asteroid S/N at 2.0~mm in the \textsc{ra13hdec-25} and \textsc{ra23hdec-25} fields.
	We base these predictions on the number of observations in which the asteroid is present in the field as well as the mean asteroid flux modeled by the STM at 2.0~mm during those observations.
	Each point in the plot represents an asteroid present in the field during at least one observation.
	The dotted lines trace out levels of constant S/N at values equal to 1, 5, and 10.}
	\label{fig:SN}
\end{figure*}

Using the methods described in Section \ref{sec:query}, we compile a list of 54 objects in \textsc{ra13hdec-25} and 82 objects in \textsc{ra13hdec-25} that are within the fields during at least one observation, and we predict the integrated S/N for each asteroid.
This information---mean model flux, number of observations present, and corresponding predicted S/N at 2.0~mm---is summarized in Figure \ref{fig:SN}.
We present these predictions only at 2.0~mm because we expect a larger emitted flux at shorter wavelengths and a better point-source sensitivity in that band.
As seen in Figure \ref{fig:SN}, only three asteroids are present for long enough and with great enough mean flux to be detected at S/N $> 5$.
We predict that the only asteroid with S/N $> 5$ in \textsc{ra13hdec-25} is (324) Bamberga with S/N $\sim 12.5$ and that the only asteroids with S/N $> 5$ in \textsc{ra23hdec-25} are (13) Egeria and (22) Kalliope with S/N $\sim 13.5$ and $\sim 10.7$, respectively.
We performed the differencing and coadding procedures described in Section \ref{sec:detect} on all 136 selected objects in our fields, and we were indeed able to detect only those 3 asteroids with high significance.
The rest of this section will focus on those 3 asteroids.

\begin{figure*}[h]
	\centering
	\includegraphics[width=0.8\linewidth]{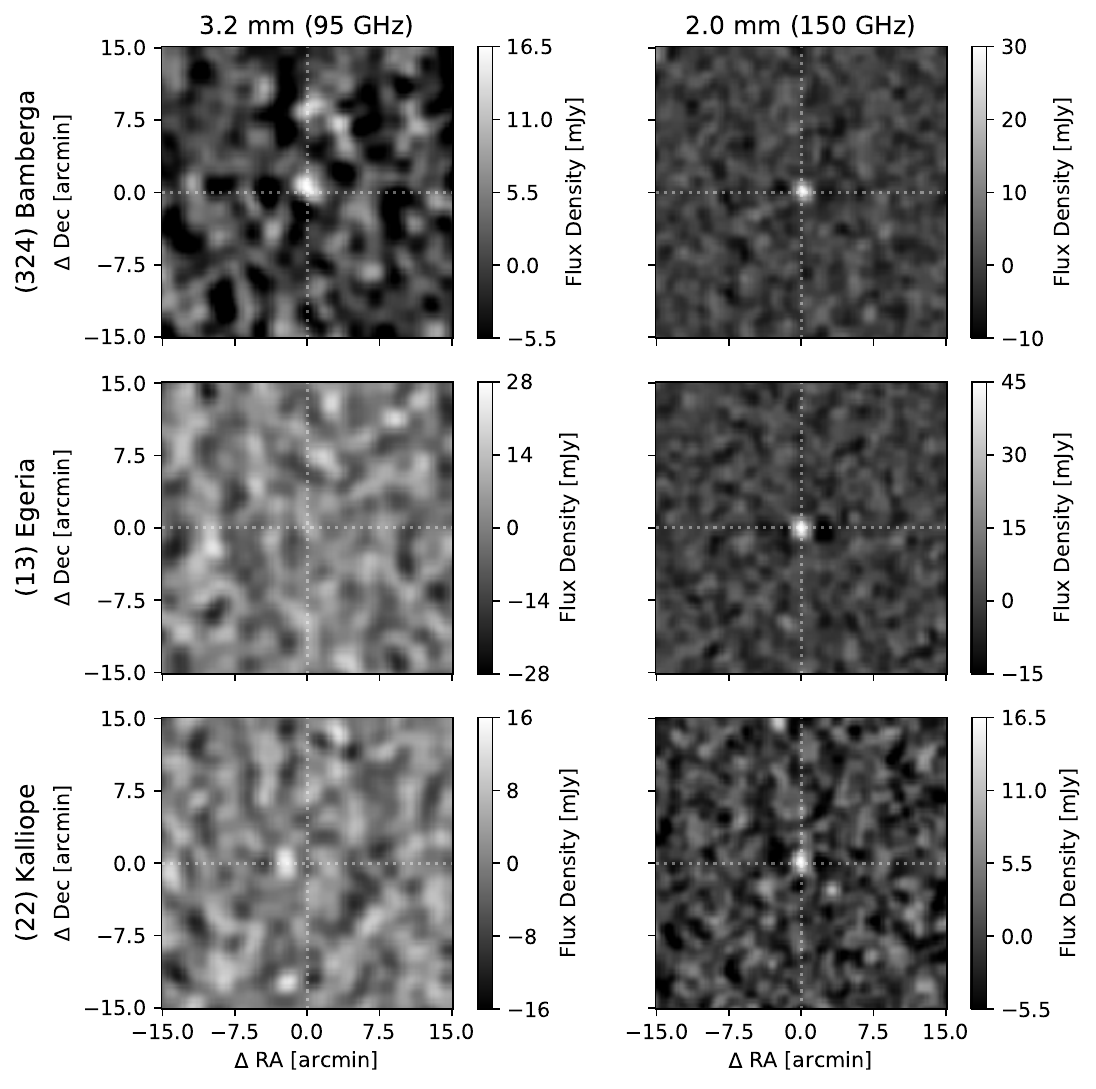}
	\caption{Mean flux measurements of (324) Bamberga (top horizontal panels), (13) Egeria (middle horizontal panels), and (22) Kalliope (bottom horizontal panels) at 3.2~mm (left vertical panels) and 2.0~mm (right vertical panels).
	Color scales for (13) Egeria and (22) Kalliope at 3.2~mm are set at 4-sigma levels; the rest peak near the mean flux values detected for each asteroid.}
	\label{fig:BEK}
\end{figure*}

\begin{figure}[h]
	\includegraphics[width=1\linewidth]{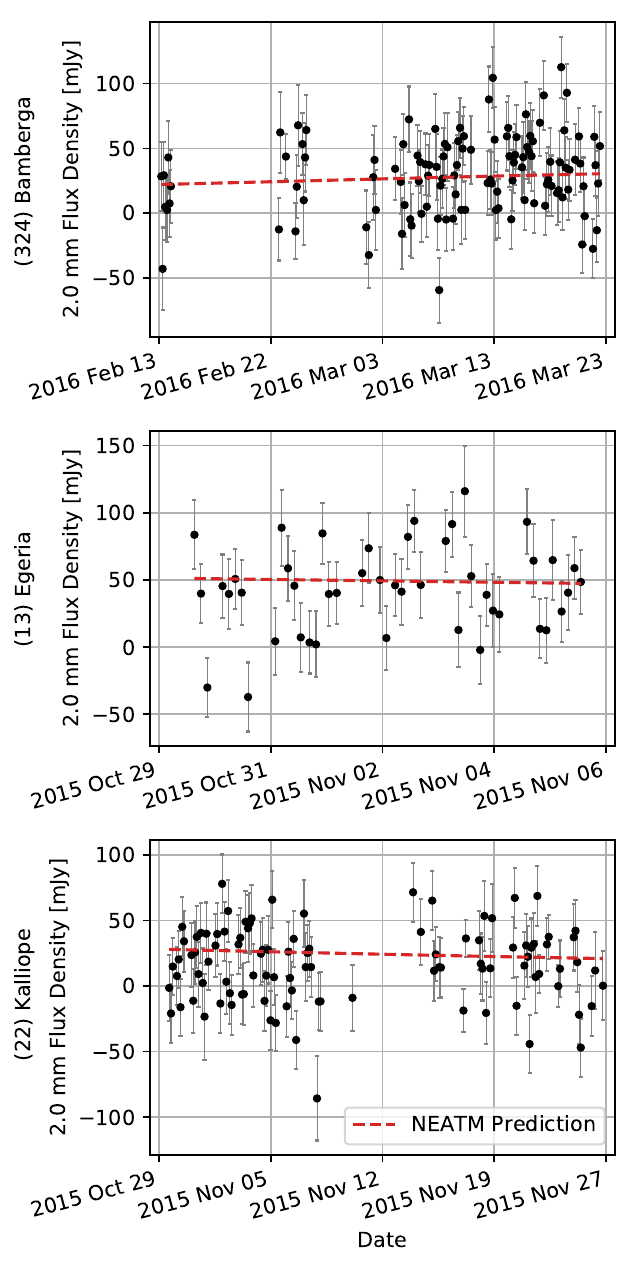}
	\caption{Light curves of (324) Bamberga (top panel), (13) Egeria (middle panel), and (22) Kalliope (bottom panel) at 2.0~mm.
	The dashed red lines represent NEATM predictions for flux density.
	(324) Bamberga is the only asteroid for which we detect statistically significant variation in the light curve of the form predicted by the NEATM.}
	\label{fig:unfolded_lightcurves}
\end{figure}

(324) Bamberga is a large MBA with an effective body diameter of 220.7~km \citep{masiero14}.
We observe (324) Bamberga in \textsc{ra13hdec-25} during 115 observations between 2016 February 13 and 2016 March 22 with mean observing geometry of 3.58~AU solar distance, 2.78~AU Earth distance, and $10.2^\circ$ solar phase angle.
Its trajectory during this time is plotted in Figure \ref{fig:trajectories}, and details of observation geometries are included in Appendix \ref{sec:obs_geoms}.
Using methods described in Section \ref{sec:detect}, we coadd observations made during this time, and we show cutouts of the resulting maps in Figure \ref{fig:BEK}.
From these maps, we detect the asteroid with S/N $=4.1$ and S/N $=11.2$ and record an average flux of $16.0 \pm 3.9$~mJy and $30.6 \pm 2.7$~mJy, at 3.2~mm and 2.0~mm, respectively, corresponding to a spectral index of $1.5 \pm 0.6$.
The measured average flux levels are roughly consistent with those predicted by the NEATM, as shown in Table \ref{tab:detections}.
We calculate expected mean flux density, effective emissivity, and brightness temperature in Table \ref{tab:detections} using 0.89 for the NEATM beaming parameter, as reported by NEOWISE \citep{mainzer19}.
Its light curve is plotted along with flux predicted by the NEATM in Figure \ref{fig:unfolded_lightcurves}.
We detect variation in the light curve of the form predicted by the NEATM with a $\chi^2$ difference of 5.08 compared to a constant flux model, corresponding to a p-value of 0.0014.

\begin{figure*}[h]
\begin{center}
	\includegraphics[width=0.8\linewidth]{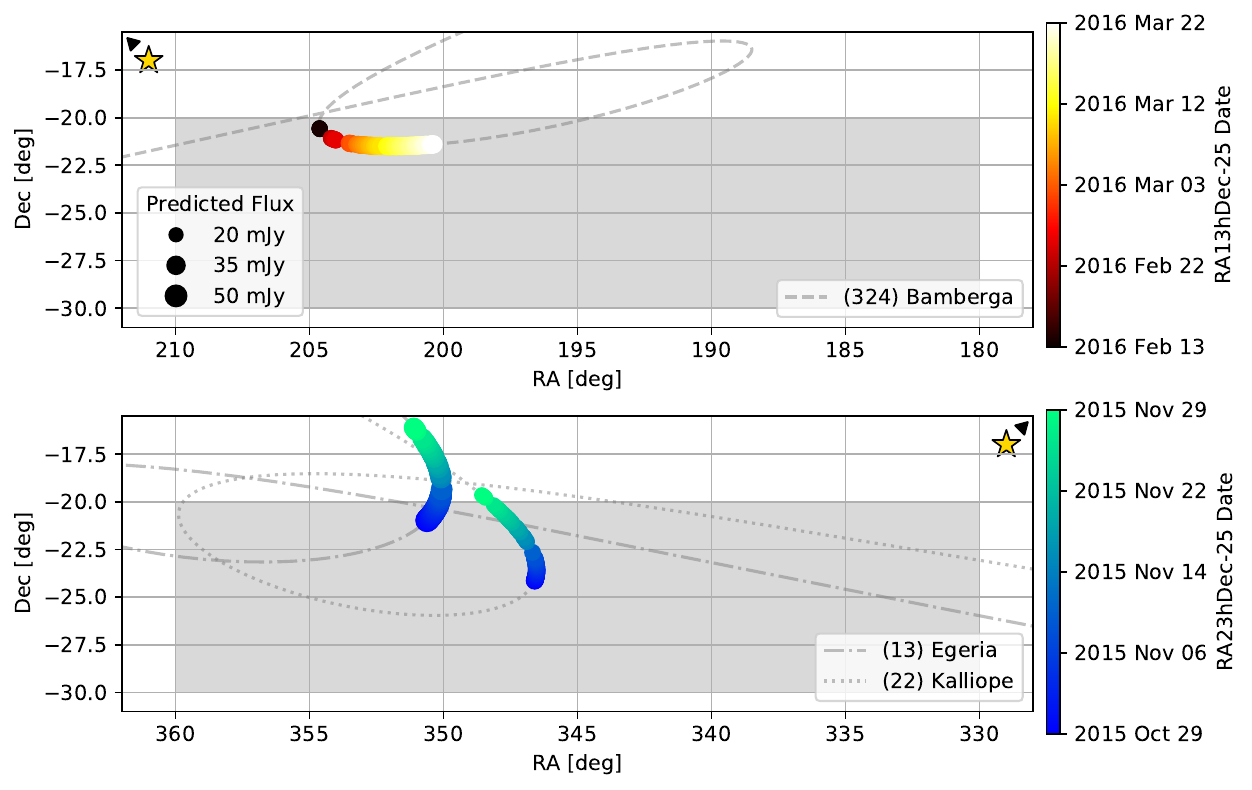}
	\caption{Trajectories of (324) Bamberga through \textsc{ra13hdec-25} (top panel) and (13) Egeria and (22) Kalliope through \textsc{ra23hdec-25} (bottom panel).
	The shaded gray regions represent the field boundaries.
	The dotted and dashed lines represent the asteroids' long-term trajectories, which trace out loops due to parallax motion.
	Each colored dot represents the asteroids' positions at the time of an observation of the field.
	The size of each dot represents flux density at 2.0~mm predicted by the NEATM.
	The stars and arrows point toward the mean locations of the sun during the observation periods.}
	\label{fig:trajectories}
\end{center}
\end{figure*}

\begin{table*}[h]
\centering
\caption{Asteroid Detections and Constrained Properties}
\begin{tabular}{lcccccc}
			             					& \multicolumn{2}{c}{(324) Bamberga} 	& \multicolumn{2}{c}{(13) Egeria} 		& \multicolumn{2}{c}{(22) Kalliope} \\
           				 					& 3.2~mm           	& 2.0~mm 			& 3.2~mm         	& 2.0~mm          	& 3.2~mm        	& 2.0 mm         \\
\hline
Measured S/N             					& 4.1              	& 11.2 				& 1.7            	& 10.4            	& -0.8          	& 6.1             \\
Measured Mean Flux (mJy) 					& 16.0 $\pm$ 3.9   	& 30.6 $\pm$ 2.7 	& 11.6 $\pm$ 6.9 	& 44.5 $\pm$ 4.3 	& -3.1 $\pm$ 4.0 	& 16.5 $\pm$ 2.7      \\
Predicted Mean Flux (mJy)    				& 11.8             	& 28.1             	& 20.7 				& 49.6            	& 10.7          	& 25.6            \\
Predicted Flux Range (mJy)					& 9.3 -- 11.9		& 22.2 -- 28.5		& 17.1 -- 21.4		& 40.8 -- 51.4		& 9.2 -- 11.2		& 21.9 -- 26.7	\\
Effective Emissivity	 					& 1.36 $\pm$ 0.33	& 1.09 $\pm$ 0.10	& $< 1.23$			& 0.90 $\pm$ 0.09 	& $< 0.46$		 	& 0.64 $\pm$ 0.11	  \\
Brightness Temperature (K AU$^{1/2}$)		& 546.9 $\pm$ 132.9	& 438.4 $\pm$ 38.4	& $< 488.1$	   		& 356.7 $\pm$ 33.5 	& $< 176.9$	 		& 246.6 $\pm$ 39.6	  \\
Max. Modulation Amplitude					& $\cdots$			& $< 33.6\%$ 		& $\cdots$		   	& $< 43.3\%$ 	 	& $\cdots$	 		& $ < 73.2\%$	\\
Spectral Index						& \multicolumn{2}{c}{1.5 $\pm$ 0.6}	& \multicolumn{2}{c}{3.1 $\pm$ 1.4}	& \multicolumn{2}{c}{$> 1.8$} \\
\hline
\end{tabular}
\begin{flushleft}
{\small
Measurements of flux density and S/N, predictions of flux density, and measurements of or limits on effective emissivity, brightness temperature as a function of solar distance, and light curve modulation amplitude for the three asteroids in \textsc{ra13hdec-25} and \textsc{ra23hdec-25} with S/N $> 5$.
We calculate predicted mean flux using NEATM, while we calculate lower and upper values of predicted flux range with FRM and STM, respectively.
We report $\pm$ values with 1-sigma significance and upper limits with 2-sigma significance.
Uncertainties reported in this table are pure statistical uncertainties and should be added in quadrature with subdominant systematic uncertainties.
The lower limit on spectral index for (22) Kalliope is estimated using upper and lower 2-sigma flux measurements at 3.2~mm and 2.0~mm, respectively.}
\end{flushleft}
\label{tab:detections}
\end{table*}

(13) Egeria is a large MBA with an effective body diameter of 202.6~km \citep{nugent15}.
We observe (13) Egeria in \textsc{ra23hdec-25} during 45 observations from 2015 October 29 until it exits the field on 2015 November 5 with mean observing geometry of 2.68~AU solar distance, 2.02~AU Earth distance, and $18.2^\circ$ solar phase angle.
Its trajectory during this time is plotted in Figure \ref{fig:trajectories}, and details of observation geometries are included in Appendix \ref{sec:obs_geoms}.
We show cutouts of the averaged observation maps in Figure \ref{fig:BEK}.
From these maps, we detect the asteroid with S/N $=1.7$ and S/N $=10.4$ and record an average flux of $11.6 \pm 6.9$~mJy and $44.5 \pm 4.3$~mJy, at 3.2~mm and 2.0~mm, respectively, corresponding to a spectral index of $3.1 \pm 1.4$.
Note that unlike (324) Bamberga and (22) Kalliope, NEOWISE was unable to fit a value for the NEATM beaming parameter for (13) Egeria, so we use their assumed value of 0.95 to calculate expected mean flux density, effective emissivity, and brightness temperature in Table \ref{tab:detections} \citep{mainzer19}.
The measured average flux level at 2.0~mm is roughly consistent with that predicted by the NEATM, as shown in Table \ref{tab:detections}.
Its light curve is plotted along with flux predicted by the NEATM in Figure \ref{fig:unfolded_lightcurves}.
Egeria's light curve shows mild preference for flux change opposite to the NEATM prediction, but this measurement is not statistically significant.

(22) Kalliope is a large MBA with an effective body diameter of 167.5~km \citep{masiero14}.
We observe (22) Kalliope in \textsc{ra23hdec-25} during 100 observations from 2015 October 29 until it exits the field on 2015 November 26 with mean observing geometry of 2.76~AU solar distance, 2.25~AU Earth distance, and $19.5^\circ$ solar phase angle.
Its trajectory during this time is plotted in Figure \ref{fig:trajectories}, and details of observation geometries are included in Appendix \ref{sec:obs_geoms}.
We show cutouts of the averaged observation maps in Figure \ref{fig:BEK}.
From these maps, we detect the asteroid with S/N $=6.1$ at 2.0~mm and record an average flux of $-3.1 \pm 4.0$~mJy and $16.5 \pm 2.7$~mJy, at 3.2~mm and 2.0~mm, respectively, corresponding to an estimated lower limit on spectral index of 1.8 at 2-sigma significance.
We calculate expected mean flux density, effective emissivity, and brightness temperature in Table \ref{tab:detections} using 1.081 for the NEATM beaming parameter, as reported by NEOWISE \citep{mainzer19}.
The measured average flux level at 2.0~mm is consistent with that predicted by the NEATM and an effective emissivity of $0.64 \pm 0.11$, as shown in Table \ref{tab:detections}.
(22) Kalliope has the lowest effective emissivity of the three asteroids, and the calculated upper limit on effective emissivity at 3.2~mm is significantly lower than that calculated at 2.0~mm.
Its light curve is plotted along with flux predicted by the NEATM in Figure \ref{fig:unfolded_lightcurves}.
The light curve does not show evidence of brightness modulation beyond a constant model.

(22) Kalliope is part of a binary system with its satellite Linus.
Studies of (22) Kalliope report that Linus is dimmer by a factor of $25 \pm 5$, so we ignore the contribution of Linus to mean flux \citep{margot03}.

When we compare the light curves plotted in Figure \ref{fig:unfolded_lightcurves} with NEATM predictions, we calculate excess variance beyond the model.
This excess variance in asteroid light curves suggests that we may be missing something in either our flux calculations or our model, but we are confident that this is not due to day-to-day calibration given our checks described in Section \ref{sec:cal}.
We perform a check by creating light curves of null off-target pixels from differenced maps, and we confirm that these light curves are statistically consistent with zero.

\begin{figure}[t]
	\includegraphics[width=1\linewidth]{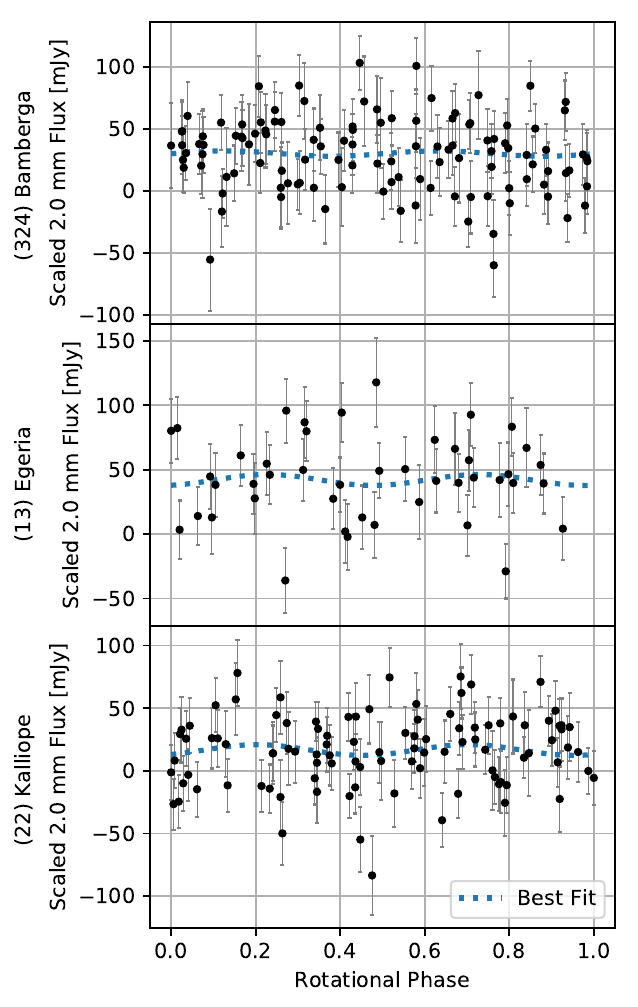}
	\caption{Phased light curves of (324) Bamberga (top panel), (13) Egeria (middle panel), and (22) Kalliope (bottom panel) at 2.0~mm, folded over rotational period and scaled to mean observed solar distance, Earth distance, and solar phase angle as described in Section \ref{sec:detect}.
	The dotted blue lines represent the sinusoidal function that best fits the data, although none of these fits are statistically significant.
	The sinusoidal periods are constrained to half the rotational periods, but the amplitudes and phases are free parameters.}
	\label{fig:folded_lightcurves}
\end{figure}

Because we detect the mean flux from these three asteroids with a high S/N at 2.0~mm, we also study their light curves at this wavelength in an attempt to detect the effect of rotation.
We calculate the Lomb-Scargle periodogram for each light curve, and we do not detect significant periodicity in any of the three.
We then adopt external constraints on the rotational period of the asteroids and attempt to detect the modulation effect in folded light curves.
We assume rotational periods based on published observations at other wavelengths reported by the Minor Planet Center's Asteroid Lightcurve Database (LCDB) \citep{warner09}.
We assume that (324) Bamberga rotates with a rotational period of 29.43 hours, (13) Egeria with a rotational period of 7.045 hours, and (22) Kalliope with a rotational period of 4.1483 hours.
We fold the light curve on the rotational periods and show the resulting phased light curves in Figure \ref{fig:folded_lightcurves}.
None of these light curves modulate enough at 2.0~mm to detect within our sensitivity, and we use that fact to set limits on maximum possible modulation amplitude, which we list in Table \ref{tab:detections}.

\section{Discussion}
\label{sec:discuss}

We measured mean flux densities for (324) Bamberga and (13) Egeria that were relatively close to NEATM predictions.
Meanwhile, we measured mean flux densities for (22) Kalliope well below the predicted values.

This discrepancy may be explained by considering the asteroids' compositions and physical properties.
For instance, radar albedo measurements can inform knowledge of near-surface density and porosity.
In general, lower radar albedo is correlated with lower density or higher porosity, and asteroids of similar classification have similar radar albedo.
(22) Kalliope, an M-type asteroid, has a radar albedo around $0.18\pm0.05$ \citep{shepard15}, consistent with an average $0.294\pm0.135$ for other M-types \citep{virkki14}.
Given (22) Kalliope's relatively high bulk density of about 3.4~$\mathrm{g} \mathrm{cm}^{-3}$ \citep{descamps08}, this suggests it has a porous surface composed of a mixture of silicates and iron.
(324) Bamberga, a C-type asteroid, and (13) Egeria, a G-type asteroid, have much lower radar albedos of $0.031\pm0.009$ and $0.059\pm0.023$, respectively \citep{magri07}.
This suggests that (324) Bamberga and (13) Egeria have lower near-surface density or higher porosity than (22) Kalliope.

Consider also the thermal inertia, $\Gamma$, of an object, defined as
\begin{equation}
	\Gamma = \sqrt{\rho C k}
\end{equation}
where $\rho$ is the object's density, $C$ is the object's heat capacity, and $k$ is the object's thermal conductivity.
Greater thermal inertia corresponds with an object's greater resistance to changing temperature.
(22) Kalliope's higher near-surface density than (324) Bamberga and (13) Egeria suggests a higher thermal inertia.
Indeed, studies have shown that M-type asteroids like (22) Kalliope in general have much higher values of thermal inertia than other carbonaceous asteroids like (324) Bamberga and (13) Egeria \citep{opeil10}.
(22) Kalliope's greater thermal inertia may impact its mm flux in two particular ways.

One effect of greater thermal inertia is that the resistance to changing temperature causes less diurnal temperature variation: the asteroid's dayside has a lower temperature and therefor decreased radiated flux, and the asteroid's nightside has a higher temperature and therefor increased radiated flux.
Overall, this effect leads to a net decrease in expected flux for most objects viewed at less than a $90^\circ$ solar phase angle, i.e. those objects with solar distance greater than 1~AU like the vast majority of objects potentially analyzed with SPT data.
Flux models like the NEATM and the STM assume the case of low thermal inertias, in which one subsolar point reaches maximum temperature equivalent to a non-rotating body.
The FRM assumes the case of large thermal inertias, wherein the asteroid's whole equator reaches maximum temperature as the asteroid rotates.
We can estimate how well we expect the NEATM to predict flux densities by calculating the ``thermal parameter'' $\Theta$ as defined by \citet{spencer89}:
\begin{equation}
	\Theta = \frac{\Gamma \sqrt{\omega}}{\epsilon \sigma T_{ss}^3}
\end{equation}
where $\omega$ is angular rotational frequency, $\epsilon$ is emissivity, $\sigma$ is the Stefan-Boltzmann constant, and $T_{ss}$ is subsolar temperature.
Small $\Theta$ values ($\sim 0$) suggest high diurnal temperature variation akin to STM and NEATM predictions, while large $\Theta$ values (roughly $\gtrsim 10$) suggest low diurnal temperature variation akin to FRM predictions.

It is tempting to explain (22) Kalliope's lower flux as caused by low diurnal temperature variation.
We can approximate emissivity around 0.9, subsolar temperature around 222~K from the NEATM, and upper limit on thermal inertia of 250~$\mathrm{J} \mathrm{m}^{-2} \mathrm{K}^{-1} \mathrm{s}^{-1/2}$ from \citet{marchis12} to find that (22) Kalliope could have $\Theta \sim 9$.
This value would be even higher assuming a lower surface temperature or emissivity.
We expect that (324) Bamberga and (13) Egeria would have a much lower thermal parameter given that we expect they have lower thermal inertia and that their rotational periods are roughly 7 and 2 times as long as (22) Kalliope's, respectively.
This calculation suggests that (22) Kalliope may have much lower diurnal temperature variation than the NEATM might predict.
However, this cannot account for the entirety of (22) Kalliope's lower effective emissivity given that our measured mean flux is significantly lower than the bound predicted by the FRM.

Furthermore, it is even clearer that rotational effects cannot solely explain (22) Kalliope's lower effective emissivity when its spin axis orientation is considered.
The Database of Asteroid Models from Inversion Techniques (DAMIT)\footnote{https://astro.troja.mff.cuni.cz/projects/damit/} is a database of three-dimensional models for many asteroids, including (22) Kalliope \citep{vdurech10}.
(22) Kalliope's proposed shape models suggest a spin axis with ecliptic latitude 3$^\circ$ and ecliptic longitude 196$^\circ$ \citep{kaasalainen02, vdurech11, hanuvs17, vernazza21}.
This low ecliptic latitude means (22) Kalliope has a large axial tilt relative to its orbital axis.
Given our viewing geometry, (22) Kalliope's subsolar point and rotational north pole are both near each other and visible to the observer.
This geometry suggests that the same areas of (22) Kalliope's surface are being heated consistently.

Another effect of greater thermal inertia is that the resistance to changing temperature means that the subsurface regions from which mm emission originates may be cooler, causing a lower measured mm flux.
Although this is a possible explanation for the observed flux from (22) Kalliope, it is also unlikely given that the same areas of (22) Kalliope's surface are being heated consistently.

Alternatively, scattering by surface particles may explain (22) Kalliope's lower measured mm flux.
For instance, (16) Psyche is another M-type asteroid which has a comparable surface and was analyzed in detail around 1.3~mm by \cite{dekleer21}.
They found (16) Psyche had a mm emissivity of $0.61\pm0.02$ which they attributed to a highly scattering surface, and they ruled out the possibility that this measurement was caused by cooler subsurface emission.
To definitively draw similar conclusions for (22) Kalliope, we would need to conduct more advanced thermophysical modeling beyond the scope of this analysis, which attempted primarily to test our ability to make these mm flux measurements.
Nevertheless, it is reassuring that our measurements of (22) Kalliope are comparable to modern measurements at mm wavelengths of other asteroids with comparable surface composition.

To interpret our limits on rotational light curve amplitude in context, we consider previous measurements of these asteroids' light curves and shapes.
Shape models for all three asteroids were recently created by \cite{vernazza21} including measures of their their elliptical-model based eccentricities $c/a$, and maximum light curve modulation is reported on the LCDB.
(324) Bamberga is fairly round with reported $c/a$ of 0.96$\pm$0.05, so we should expect a small value for modulation amplitude.
Indeed, maximum flux modulation amplitude reported on the LCDB is 12$\%$.
(13) Egeria is notably more irregular in shape and has a reported $c/a$ of 0.76$\pm$0.06.
Maximum flux modulation amplitude reported on the LCDB is as great as 54$\%$, though some reports are lower than that.
Many of these observations also show a notably asymmetric light curve, which hurts our assumption that expected folded light curve modulation will be purely sinusoidal.
Nevertheless, a pure sinusoid should still match the modulation to first order.
(22) Kalliope is the most elliptical of the three asteroids with reported $c/a$ of 0.59$\pm$0.02, so depending on viewing geometry we could expect the largest value for modulation amplitude from this object.
Maximum flux modulation amplitudes reported on the LCDB are as great as 63$\%$, though these also vary.
However, given the orientation of (22) Kalliope's spin axis as explained before, we should expect to measure a very small modulation amplitude.
Even without considering spin axis orientation for any of our three measured asteroids, our limits on modulation amplitude are less restrictive those previously determined by others.

\section{Predicted Asteroid Detections For Other SPT Surveys}
\label{sec:pred}

\begin{figure*}[h!]
	\centering
	\includegraphics[width=0.8\linewidth]{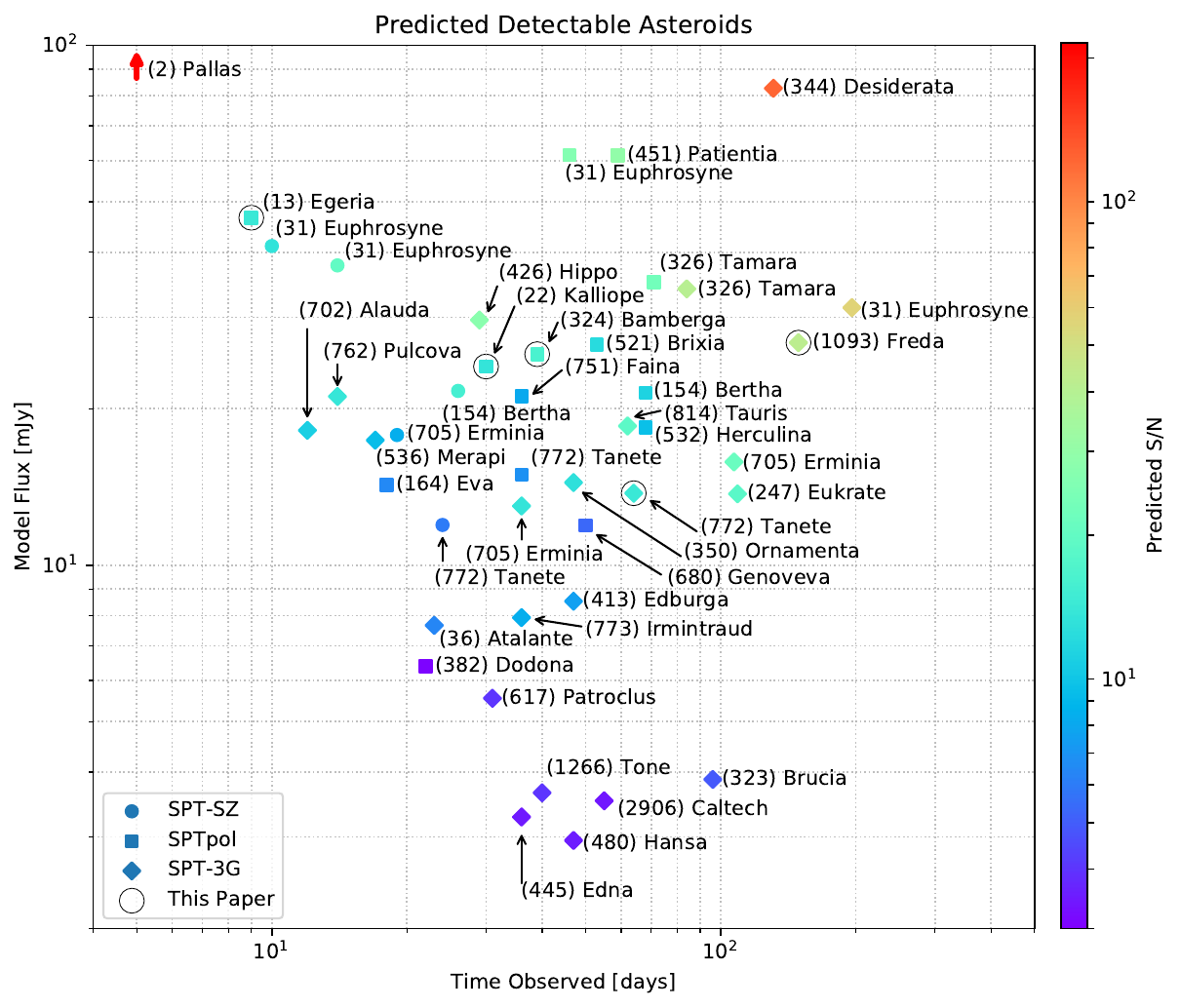}
	\caption{Objects with predicted S/N $>3$ at 2.0~mm in all historic and planned future SPT data.
	We expect to observe (2) Pallas, plotted off scale, with a mean flux density near 650~mJy.
	}
	\label{fig:pred}
\end{figure*}

As described in Section \ref{sec:telcam}, the SPT has been equipped with three separate cameras over its lifetime: SPT-SZ, SPTpol, and SPT-3G.
With SPT-SZ, observations were made of a patchwork of many fields to comprise its survey.
With SPTpol and SPT-3G, observations were made of main, deep fields during austral winters and various other fields for shorter durations during austral summers.
We refer to all observations made to date collectively as the SPT's historic data, while we refer to future observations planned with the SPT-3G camera as the SPT's future data.
Using the methods described in Section \ref{sec:query}, we compile lists of all asteroids present in historic and future data and predict their S/N.
We show all objects we expect to observe in Figure \ref{fig:pred} and summarize these predictions in Table \ref{tab:pred}.
We provide detailed field boundaries, observation periods, and object lists in Appendix \ref{sec:pred_detailed}.

We note that the SPT-3G camera has higher instantaneous sensitivity than the SPT-SZ and SPTpol cameras, so we expect to measure more asteroids and at improved sensitivities in SPT-3G data.
We expect that the improved sensitivity will allow us to place tighter constraints on both mean flux measurements and modulation amplitudes of folded light curves.
For a rough estimation, our ability to constrain modulation amplitudes will depend on the S/N available in roughly one fourth of an object's rotation period, which is roughly one half of the S/N shown in Figure \ref{fig:pred}.

\begin{table}[t]
\centering
\caption{Summary of Predicted Detections}
\begin{tabular}{lr}
		& Objects with  \\ 
Survey	& Predicted S/N \textgreater{} 3 \\
\hline
SPT-SZ							& 4 \\
SPTpol Deep (``500d'') Field	& 2 \\
SPTpol Summer Fields			& 12 \\
SPT-3G Deep (``1500d'') Field	& 14 \\
SPT-3G Summer Fields			& 10 \\
\hline
\end{tabular}
\begin{flushleft}
{\small
Summary of estimated number of detectable asteroids at 2.0~mm in completed historical and planned future surveys using the SPT.
The SPTpol and SPT-3G ``Deep Fields'' are those main fields observed during austral winters, while the ``Summer Fields'' are those other fields observed during austral summers.
Details included in Appendix \ref{sec:pred_detailed}.}
\end{flushleft}
\label{tab:pred}
\end{table}

\subsection{Validation of Prediction Model with SPT-3G 1500d Data}

As shown in Appendix \ref{sec:pred_detailed}, we should see two asteroids at high S/N in the SPT-3G main survey field during the 2020 austral winter. 
Since maps were generated following each observation of the SPT-3G survey field as part of an online data-quality monitoring pipeline, it was straightforward to extract thumbnail maps around the locations of the asteroids using methods similar to those in Section \ref{sec:detect}. 
We constructed a proof-of-concept pipeline for SPT-3G, and although the pipeline was separate, it was similar in construction and methodology to the pipeline used for the detections in SPTpol data. 

\begin{figure*}[ht!]
\centering
  \includegraphics[width=0.8\linewidth]{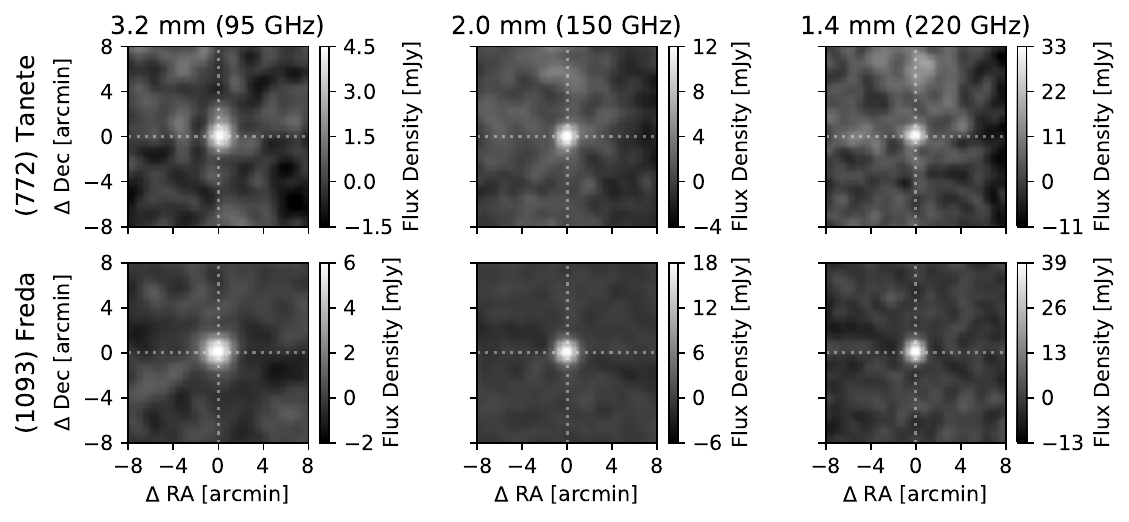}
  \caption{Mean flux measurements of (772) Tanete (top horizontal panels) and (1093) Freda (bottom horizontal panels) at 3.2~mm (left vertical panels), 2.0~mm (middle vertical panels), and 1.4~mm (right vertical panels).
  These measurements were taken with the SPT-3G camera in the main survey field during the 2020 austral winter.
  }
  \label{fig:TF}
\end{figure*}

We report the mean fluxes and signal-to-noise ratios for the two asteroids with predicted S/N $> 5$ at 2.0~mm in the 1500 deg$^2$ survey region during the austral winter 2020. 
We observe (1093) Freda in 297 observations and measure a mean flux density of $6.3\pm0.6$, $18.7\pm0.8$, and $43.0\pm 3.3$ mJy at 3.2, 2.0, and 1.4~mm respectively, corresponding to a S/N of 10.6, 22.9, and 13.1. 
We observe (772) Tanete in 156 observations and measure a mean flux density of $4.4\pm0.7$, $10.7\pm1.9$, and $27.1\pm 6.0$ mJy at 3.2, 2.0, and 1.4~mm respectively, corresponding to a S/N of 6.5, 5.6, and 4.5.
We show coadded observation maps of these asteroids in Figure \ref{fig:TF}.

\section{Conclusion}
\label{sec:conc}

In this work, we have demonstrated that we are able to detect asteroids in mm observations of the sky made with the SPT, and we show that we will detect even more asteroids in other historic and future data from the SPT.
Using maps from repeated observations of the same area of the sky over the course of months, we measure three asteroids, (324) Bamberga, (13) Egeria, and (22) Kalliope, at wavelengths of 3.2~mm and 2.0~mm with the SPTpol camera, and we report measurements of the asteroids' mean fluxes at 2.0~mm with a S/N of 11.2, 10.5, and 6.1, respectively.
We also report measurements of (324) Bamberga at 3.2~mm with a S/N of 4.1.
Although others have studied asteroid thermal emission at mm wavelengths, this work is the first focused analysis of asteroid flux using data taken with the primary science goal of characterizing the CMB.

Observing asteroids with CMB survey data expands the breadth of two separate fields of astronomy.
CMB survey scientists can now include asteroid science as part of their data analysis, and they have more scientific use for their historic data.
They may perform more focused studies of asteroids in the future, potentially including near-Earth asteroids that pass through survey fields.
Meanwhile, asteroid scientists now have access to more data on the thermal emission of asteroids.
They can make use of measurements in mm wavelengths made using CMB experiments, especially as the instantaneous sensitivity of CMB cameras improves and allows more precise time-domain astronomy.

Our measurements in mm wavelengths come at an important time for asteroid scientists, when studies like \citet{keihm13} suggest a paradigm shift in the understanding of asteroid regolith temperatures.
We measured the flux from (324) Bamberga to moderate significance at 3.2~mm and high significance at 2.0~mm and found flux densities consistent with model predictions.
We measured the flux from (13) Egeria to high significance at 2.0~mm, and we used the lack of a detection at 3.2~mm to place limits on its brightness temperature and effective emissivity at this wavelength.
We measured the flux from (22) Kalliope to moderate significance at 2.0~mm and showed a significant decrease in mm flux at 3.2~mm compared to 2.0~mm, consistent with previous studies of other large MBAs suggesting a decrease in flux at longer wavelengths.
Our measurements will help place limits on the thermal properties and composition of these asteroids' regoliths.

With historic and future data, we expect to observe 34 total asteroids, including very precise measurements of (2) Pallas and (344) Desiderata; multiple measurements of (31) Euphrosyne, (154) Bertha, (326) Tamara, (705) Erminia, and (772) Tanete; and measurements of (617) Patroclus, a target of NASA's \textit{Lucy} mission \citep{levison21}.

Using data from SPTpol, we made significant measurements of (13) Egeria and (22) Kalliope at only one wavelength, and we did not have high enough sensitivity in individual observations to describe light curves to high accuracy.
However, SPT-3G will improve on both of these limitations with its higher sensitivity and third wavelength band, as shown by the detections of (772) Tanete and (1093) Freda.
SPT-3G's higher sensitivity will also allow us to observe fainter objects than with SPTpol.
In fact, SPT-3G and other future CMB surveys may have to mask asteroids during transient source analysis, since as of the writing of this paper, objects like (344) Desiderata are bright enough to trigger the SPT-3G transient alert system.
Furthermore, many of the asteroids detectable with SPT-3G will be observed for longer periods of time than those with SPT-SZ or SPTpol.
The longer observation times and higher instantaneous sensitivity of SPT-3G will allow us to place tighter constraints on modulation amplitude for those asteroids, and potentially to detect modulations of more than a few percent.
With SPT-3G sensitivity, we will be able to observe more asteroids at more wavelengths, and, for many of them, over longer times.

By repurposing historic and future data from the SPT, we will observe asteroids without needing to dedicate telescope observation time to do so.
The measurements we make will provide new constraints on the thermal properties and compositions of asteroid regoliths.

\section{Acknowledgements}

The South Pole Telescope program is supported by the National Science Foundation (NSF) through award OPP-1852617.

This paper makes use of the following ALMA data: ADS/JAO.ALMA\#2011.0.00001.CAL. ALMA is a partnership of ESO (representing its member states), NSF (USA) and NINS (Japan), together with NRC (Canada), MOST and ASIAA (Taiwan), and KASI (Republic of Korea), in cooperation with the Republic of Chile. The Joint ALMA Observatory is operated by ESO, AUI/NRAO and NAOJ.

This publication also makes use of data products from NEOWISE, which is a project of the Jet Propulsion Laboratory/California Institute of Technology, funded by the Planetary Science Division of the National Aeronautics and Space Administration.

MA and JV acknowledge support from the Center for AstroPhysical Surveys at the National Center for Supercomputing Applications in Urbana, IL.
JV acknowledges support from the Sloan Foundation.

The authors thank Jeff McMahon\footnote{https://orcid.org/0000-0002-7245-4541} for support in manuscript preparation through the Science Writing Practicum taught as part of U.~Chicago's ``Data Science in Energy and Environmental Research'' NRT training program, NSF grant \#DGE-1735359.

\bibliography{asteroids21}

\appendix

\section{Observation Geometries}
\label{sec:obs_geoms}

\begin{figure*}[h!]
	\centering
	\includegraphics[width=0.6\linewidth]{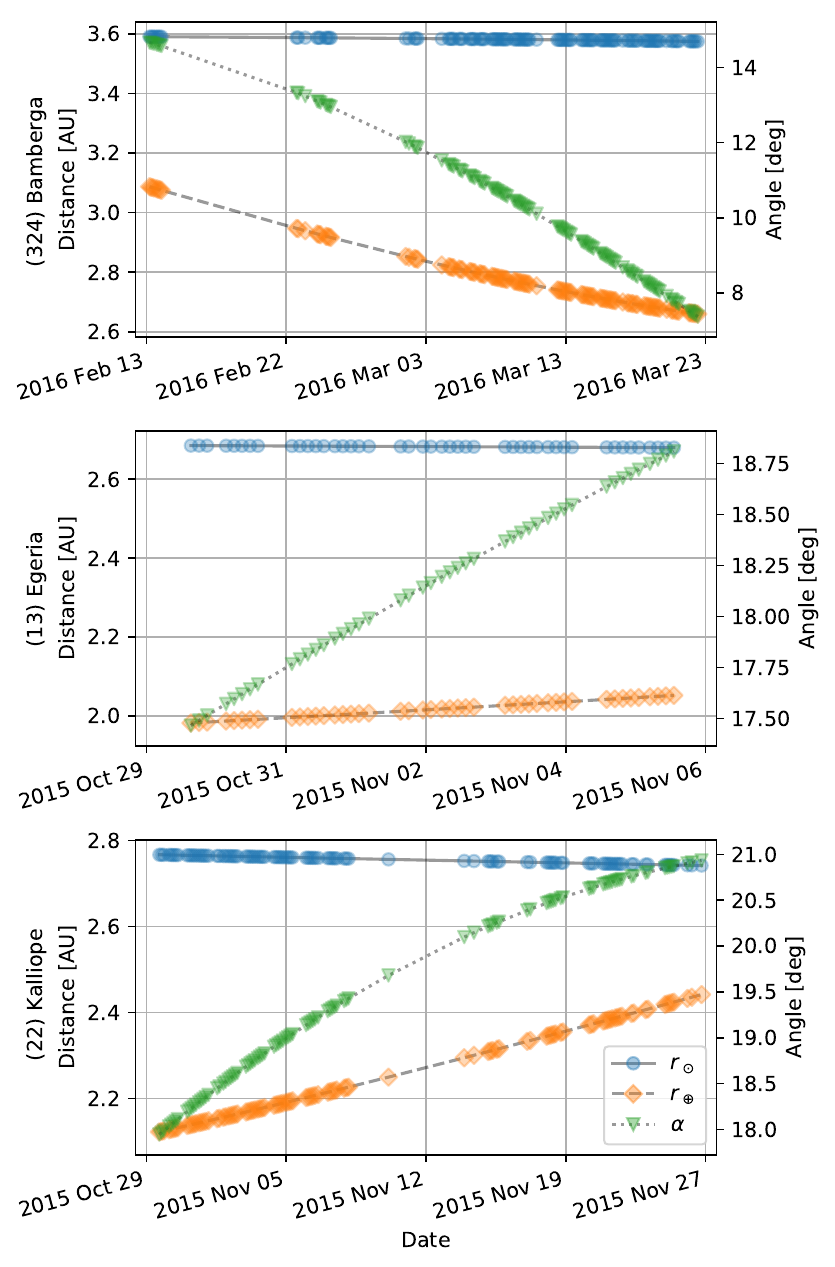}
	\caption{Observation geometries for (324) Bamberga (top panel), (13) Egeria (middle panel), and (22) Kalliope (bottom panel).  Each point represents one observation.  Solar distance $r_\odot$ and Earth distance $r_\oplus$ are plotted in AU, and solar phase angle $\alpha$ is plotted in degrees.}
	\label{fig:obs_geoms}
\end{figure*}

This section describes details of the observation geometries for the 3 asteroids analyzed in Section \ref{sec:results}.  Figure \ref{fig:obs_geoms} shows the asteroids' solar distance, earth distance, and solar phase angle during observations.  This information was queried from JPL HORIZONS.

\newpage
\section{Asteroids in Historic Data}
\label{sec:pred_detailed}

This section contains a detailed definition of each observation field's boundaries, time range, mean solar R.A., and objects present for all historic and planned future observations using the SPT-SZ, SPTpol, and SPT-3G cameras, constructed using the methods described in Section \ref{sec:query}.
We plot the results in Figure \ref{fig:pred} and list details in Tables \ref{tab:pred_SZ}, \ref{tab:pred_pol}, and \ref{tab:pred_3G}, respectively.

\setlength\LTcapwidth{\textwidth}   
\setlength\LTleft{0pt}
\setlength\LTright{0pt} 

\begin{longtable}{@{\extracolsep{\fill}}p{4.2cm}cccccc@{}}
\caption{Detections at 2.0~mm Predicted with SPT-SZ} \\
			&					&					&				& Mean Solar 		& Objects with 						& Predicted \\
Field Name	& R.A.~($^\circ$)	& Decl.~($^\circ$)	& Time Range	& R.A.~($^\circ$) 	& Predicted S/N \textgreater{} 3	& S/N \\ \hline
\textsc{ra5h30dec-55}	& [75, 90] 		& [-60, -50] 	& 2008 Feb 13 -- 2008 Jun 05 	& 19 	& $\cdots$	& $\cdots$ \\
\textsc{ra5h30dec-55}	& [75, 90] 		& [-60, -50] 	& 2011 Jan 13 -- 2011 Feb 20 	& 314 	& $\cdots$	& $\cdots$ \\
\textsc{ra5h30dec-55}	& [75, 90] 		& [-60, -50] 	& 2011 Aug 23 -- 2011 Aug 24 	& 152 	& $\cdots$	& $\cdots$ \\
\textsc{ra5h30dec-55}	& [75, 90] 		& [-60, -50] 	& 2011 Sep 22 -- 2011 Oct 04 	& 184 	& $\cdots$	& $\cdots$ \\
\textsc{ra5h30dec-55}	& [75, 90] 		& [-60, -50] 	& 2011 Oct 20 -- 2011 Nov 13 	& 216 	& $\cdots$	& $\cdots$ \\
\textsc{ra23h30dec-55}	& [345, 360] 	& [-60, -50] 	& 2008 May 27 -- 2008 Sep 25 	& 125 	& $\cdots$	& $\cdots$ \\
\textsc{ra23h30dec-55}	& [345, 360] 	& [-60, -50] 	& 2010 Apr 15 -- 2010 May 13 	& 36 	& $\cdots$	& $\cdots$ \\
\textsc{ra21hdec-60}	& [300, 330] 	& [-65, -55] 	& 2009 Jan 31 -- 2009 Jul 01 	& 26 	& $\cdots$	& $\cdots$ \\
\textsc{ra3h30dec-60}	& [30, 75] 		& [-65, -55] 	& 2009 Feb 04 -- 2009 Mar 30 	& 344 	& $\cdots$	& $\cdots$ \\
\textsc{ra21hdec-50}	& [300, 330]	& [-55, -45] 	& 2009 Jul 23 -- 2009 Aug 10 	& 131 	& (705) Erminia	& $8.4$ \\
\textsc{ra21hdec-50}	& [300, 330]	& [-55, -45] 	& 2009 Sep 01 -- 2009 Nov 10 	& 192 	& $\cdots$	& $\cdots$ \\
\textsc{ra4h10dec-50}	& [50, 75]		& [-55, -45] 	& 2010 Feb 03 -- 2010 Apr 13 	& 349 	& $\cdots$	& $\cdots$ \\
\textsc{ra0h50dec-50}	& [0, 25]		& [-55, -45] 	& 2010 May 13 -- 2010 Jun 18 	& 68 	& $\cdots$	& $\cdots$ \\
\textsc{ra2h30dec-50}	& [25, 50]		& [-55, -45] 	& 2010 Jun 19 -- 2010 Jul 28 	& 107 	& $\cdots$	& $\cdots$ \\
\textsc{ra1hdec-60}		& [0, 30]		& [-65, -55] 	& 2010 Jul 29 -- 2010 Sep 11 	& 149 	& $\cdots$	& $\cdots$ \\
\textsc{ra5h30dec-45}	& [75, 90]		& [-50, -40] 	& 2010 Oct 07 -- 2010 Nov 05 	& 206 	& $\cdots$	& $\cdots$ \\
\textsc{ra6h30dec-55}	& [90, 105]		& [-60, -50] 	& 2010 Oct 07 -- 2010 Nov 13 	& 210 	& $\cdots$	& $\cdots$ \\
\textsc{ra6h30dec-55}	& [90, 105]		& [-60, -50] 	& 2011 Mar 09 -- 2011 Mar 23 	& 355 	& $\cdots$	& $\cdots$ \\
\textsc{ra6h30dec-55}	& [90, 105]		& [-60, -50] 	& 2011 Jul 15 -- 2011 Jul 17 	& 115 	& $\cdots$	& $\cdots$ \\
\textsc{ra23hdec-62.5}	& [330, 360]	& [-65, -60] 	& 2010 Sep 12 -- 2010 Oct 06 	& 181 	& $\cdots$	& $\cdots$ \\
\textsc{ra23hdec-62.5}	& [330, 360]	& [-65, -60] 	& 2010 Apr 24 -- 2010 Jul 15 	& 72 	& $\cdots$	& $\cdots$ \\
\textsc{ra21hdec-42.5}	& [300, 330]	& [-45, -40] 	& 2010 Sep 12 -- 2010 Oct 07 	& 181 	& (31) Euphrosyne	& $20.1$ \\
						&				&				&								&		& (154) Bertha	& $15.7$ \\
\textsc{ra21hdec-42.5}	& [300, 330]	& [-45, -40] 	& 2010 Apr 21 -- 2010 Jul 14 	& 70 	& $\cdots$	& $\cdots$ \\
\textsc{ra22h30dec-55}	& [330, 345]	& [-60, -50] 	& 2010 Sep 12 -- 2010 Oct 05 	& 180 	& $\cdots$	& $\cdots$ \\
\textsc{ra22h30dec-55}	& [330, 345]	& [-60, -50] 	& 2010 Apr 11 -- 2010 Apr 21 	& 24 	& $\cdots$	& $\cdots$ \\
\textsc{ra22h30dec-55}	& [330, 345]	& [-60, -50] 	& 2010 May 13 -- 2010 Jul 10 	& 79 	& $\cdots$	& $\cdots$ \\
\textsc{ra23hdec-45}	& [330, 360]	& [-50, -40] 	& 2010 Sep 12 -- 2010 Oct 05 	& 180 	& (31) Euphrosyne	& $13.3$ \\
						&				&				&								&		& (772) Tanete	& $6.0$ \\
\textsc{ra23hdec-45}	& [330, 360]	& [-50, -40] 	& 2011 Mar 24 -- 2011 Apr 11 	& 11 	& $\cdots$	& $\cdots$ \\
\textsc{ra23hdec-45}	& [330, 360]	& [-50, -40] 	& 2011 May 13 -- 2011 Jul 17 	& 82 	& $\cdots$	& $\cdots$ \\
\textsc{ra6hdec-62.5}	& [75, 105]		& [-65, -60] 	& 2010 Sep 12 -- 2010 Oct 07 	& 181 	& $\cdots$	& $\cdots$ \\
\textsc{ra6hdec-62.5}	& [75, 105]		& [-65, -60] 	& 2011 Jan 11 -- 2011 Feb 28 	& 317 	& $\cdots$	& $\cdots$ \\
\textsc{ra3h30dec-42.5}	& [30, 75]		& [-45, -40] 	& 2010 Sep 12 -- 2010 Oct 09 	& 182 	& $\cdots$	& $\cdots$ \\
\textsc{ra3h30dec-42.5}	& [30, 75]		& [-45, -40] 	& 2011 Mar 01 -- 2011 Mar 09 	& 345 	& $\cdots$	& $\cdots$ \\
\textsc{ra3h30dec-42.5}	& [30, 75]		& [-45, -40] 	& 2011 Jul 17 -- 2011 Aug 27 	& 136 	& $\cdots$	& $\cdots$ \\
\textsc{ra1hdec-42.5}	& [0, 30]		& [-45, -40] 	& 2010 Sep 12 -- 2010 Oct 06 	& 181 	& $\cdots$	& $\cdots$ \\
\textsc{ra1hdec-42.5}	& [0, 30]		& [-45, -40] 	& 2011 Aug 28 -- 2011 Sep 19 	& 166 	& $\cdots$	& $\cdots$ \\
\textsc{ra1hdec-42.5}	& [0, 30]		& [-45, -40] 	& 2011 Oct 05 -- 2011 Oct 08 	& 192 	& $\cdots$	& $\cdots$ \\
\textsc{ra6h30dec-45}	& [90, 105]		& [-50, -40] 	& 2010 Sep 12 -- 2010 Oct 03 	& 179 	& $\cdots$	& $\cdots$ \\
\textsc{ra6h30dec-45}	& [90, 105]		& [-50, -40] 	& 2011 Sep 19 -- 2011 Oct 28 	& 194 	& $\cdots$	& $\cdots$ \\ \hline
\label{tab:pred_SZ}
\end{longtable}

\begin{longtable}{@{\extracolsep{\fill}}p{4.2cm}cccccc@{}}
\caption{Detections at 2.0~mm Predicted with SPTpol} \\
			&					&					&				& Mean Solar 		& Objects with 						& Predicted \\
Field Name	& R.A.~($^\circ$)	& Decl.~($^\circ$)	& Time Range	& R.A.~($^\circ$) 	& Predicted S/N \textgreater{} 3	& S/N \\ \hline
\textsc{ra23h30dec-55} (``100d'')	& [345, 360]	& [-60, -50]	& 2012 Feb 17 -- 2012 Nov 21 	& 103 	& $\cdots$	& $\cdots$ \\ 
\textsc{ra23h30dec-55} (``100d'')	& [345, 360]	& [-60, -50]	& 2013 Feb 08 -- 2013 Apr 30 	& 0 	& $\cdots$	& $\cdots$ \\ 
\textsc{ra0hdec-57.5} (``500d'')	& [-30, 30]		& [-65, -50]	& 2013 Apr 30 -- 2013 Nov 27 	& 141 	& $\cdots$	& $\cdots$ \\
\textsc{ra0hdec-57.5} (``500d'')	& [-30, 30]		& [-65, -50]	& 2014 Mar 25 -- 2014 Dec 12 	& 130 	& $\cdots$	& $\cdots$ \\
\textsc{ra0hdec-57.5} (``500d'')	& [-30, 30]		& [-65, -50]	& 2015 Mar 27 -- 2015 Oct 26 	& 109 	& (772) Tanete	& 6.9 \\
\textsc{ra0hdec-57.5} (``500d'')	& [-30, 30]		& [-65, -50]	& 2016 Mar 23 -- 2016 Sep 08 	& 84 	& (326) Tamara	& 35.0 \\
\textsc{ra0p75hdec-31} (``KiDS'')	& [-30, 52.5]	& [-36, -26]	& 2016 Sep 09 -- 2016 Nov 15 	& 198 	& (31) Euphrosyne	& 26.5 \\
									&				&				&								&		& (154) Bertha	& 11.3 \\
									&				&				&								&		& (451) Patientia	& 30.0 \\
									&				&				&								&		& (521) Brixia	& 12.3 \\
									&				&				&								&		& (532) Herculina	& 9.6 \\
									&				&				&								&		& (680) Genoveva	& 5.4 \\
									&				&				&								&		& (751) Faina	& 8.1 \\
\textsc{ra1hdec-25}		& [0, 30]		& [-30, -20]	& 2015 Dec 01 -- 2016 Feb 01 	& 280 	& $\cdots$	& $\cdots$ \\
\textsc{ra1hdec-35}		& [0, 30]		& [-40, -30]	& 2014 Jan 12 -- 2014 Feb 04 	& 305 	& $\cdots$	& $\cdots$ \\
\textsc{ra1hdec-35}		& [0, 30]		& [-40, -30]	& 2015 Dec 22 -- 2015 Dec 23 	& 270 	& $\cdots$	& $\cdots$ \\
\textsc{ra3hdec-25}		& [30, 60]		& [-30, -20]	& 2014 Feb 22 -- 2014 Mar 24 	& 349 	& $\cdots$	& $\cdots$ \\
\textsc{ra3hdec-25}		& [30, 60]		& [-30, -20]	& 2015 Feb 18 -- 2015 Feb 27 	& 335 	& $\cdots$	& $\cdots$ \\
\textsc{ra3hdec-35}		& [30, 60]		& [-40, -30]	& 2014 Feb 04 -- 2014 Feb 16 	& 323 	& $\cdots$	& $\cdots$ \\
\textsc{ra3hdec-35}		& [30, 60]		& [-40, -30]	& 2015 Feb 03 -- 2015 Feb 23 	& 326 	& $\cdots$	& $\cdots$ \\
\textsc{ra5hdec-25}		& [60, 90]		& [-30, -20]	& 2015 Mar 01 -- 2015 Mar 26 	& 353 	& $\cdots$	& $\cdots$ \\
\textsc{ra5hdec-35}		& [60, 90]		& [-40, -30]	& 2014 Feb 17 -- 2014 Mar 07 	& 339 	& $\cdots$	& $\cdots$ \\
\textsc{ra5hdec-35}		& [60, 90]		& [-40, -30]	& 2015 Jan 22 -- 2015 Jan 22 	& 304 	& $\cdots$	& $\cdots$ \\
\textsc{ra11hdec-25}	& [150, 180]	& [-30, -20]	& 2016 Jan 23 -- 2016 Feb 12 	& 315 	& $\cdots$	& $\cdots$ \\
\textsc{ra11hdec-25}	& [150, 180]	& [-30, -20]	& 2016 Mar 07 -- 2016 Mar 07 	& 348 	& $\cdots$	& $\cdots$ \\
\textsc{ra13hdec-25}	& [180, 210]	& [-30, -20]	& 2016 Feb 13 -- 2016 Mar 22 	& 344 	& (324) Bamberga$^*$	& 16.3 \\
						&				&				&								& 		& (382) Dodona			& 3.1 \\
\textsc{ra23hdec-25}	& [330, 360]	& [-30, -20]	& 2015 Oct 29 -- 2015 Nov 29 	& 255 	& (13) Egeria$^*$		& 14.2 \\
						&				&				&								& 		& (22) Kalliope$^*$		& 13.2 \\
\textsc{ra23hdec-35}	& [330, 360]	& [-40, -30]	& 2013 Nov 27 -- 2014 Jan 11 	& 267 	& (164) Eva				& 6.5 \\
\textsc{ra23hdec-35}	& [330, 360]	& [-40, -30]	& 2015 Jan 26 -- 2015 Feb 03 	& 312 	& $\cdots$	& $\cdots$ \\ \hline
\multicolumn{5}{l}{\footnotesize{$*$ Focus of this paper's analysis.}} \\
\label{tab:pred_pol}
\end{longtable}

\begin{longtable}{@{\extracolsep{\fill}}p{4.5cm}cccccc@{}}
\caption{Detections at 2.0~mm Predicted with SPT-3G} \\
			&					&					&				& Mean Solar 		& Objects with 						& Predicted \\
Field Name	& R.A.~($^\circ$)	& Decl.~($^\circ$)	& Time Range	& R.A.~($^\circ$) 	& Predicted S/N \textgreater{} 3	& S/N \\ \hline
\textsc{ra0hdec-56} (``1500d'')						& [-50, 50] & [-70, -42]	& 2019 Mar 21 -- 2019 Dec 30 	& 137 	& (413) Edburga		& 7.6 \\
													&			&				&								&		& (705) Erminia		& 21.3\\
\textsc{ra0hdec-56} (``1500d'')						& [-50, 50] & [-70, -42]	& 2020 Mar 23 -- 2020 Nov 25 	& 122 	& (772) Tanete		& 14.3 \\
													&			&				&								&		& (1093) Freda		& 42.6 \\
													&			&				&								&		& (2906) Caltech	& 3.4 \\
\textsc{ra0hdec-56} (``1500d'')						& [-50, 50] & [-70, -42]	& 2021 Mar 01 -- 2021 Dec 01 	& 114 	& (31) Euphrosyne	& 57.0 \\
													&			&				&								&		& (344) Desiderata	& 123.1 \\
													&			&				&								&		& (814) Tauris		& 19.0 \\
\textsc{ra0hdec-56} (``1500d'')$^{\dagger}$			& [-50, 50] & [-70, -42] 	& 2022 Mar 22 -- 2022 Nov 30 	& 123 	& $\cdots$			& $\cdots$ \\
\textsc{ra0hdec-56} (``1500d'')$^{\dagger}$			& [-50, 50] & [-70, -42] 	& 2023 Mar 22 -- 2023 Nov 30 	& 123 	& (247) Eukrate		& 18.6 \\
													&			&				&								&		& (323) Brucia		& 4.9 \\
													&			&				&								&		& (326) Tamara		& 40.6 \\
													&			&				&								&		& (350) Ornamenta	& 12.9 \\
													&			&				&								&		& (536) Merapi		& 9.3 \\
													&			&				&								&		& (617) Patroclus	& 4.0 \\
\textsc{ra5hdec-45.5} (``Western'')					& [50, 100]	& [-63, -28]	& 2020 Feb 10 -- 2020 Mar 22 	& 342 	& $\cdots$	& $\cdots$ \\
\textsc{ra5hdec-45.5} (``Western'')$^{\ddagger}$	& [50, 100]	& [-63, -28] 	& 2021 Jan 12 -- 2021 Feb 02 	& 305 	& $\cdots$	& $\cdots$ \\
\textsc{ra5hdec-45.5} (``Western'')					& [50, 100]	& [-63, -28]	& 2021 Dec 25 -- 2022 Feb 13 	& 300 	& $\cdots$	& $\cdots$ \\
\textsc{ra5hdec-45.5} (``Western'')$^{\dagger}$		& [50, 100]	& [-63, -28]	& 2022 Dec 25 -- 2023 Feb 13 	& 300 	& (2) Pallas	& 214.6 \\
\textsc{ra1h40dec-35} (``Mid-North'')				& [0, 50] 	& [-42, -28] 	& 2020 Dec 01 -- 2021 Jan 21 	& 275 	& $\cdots$	& $\cdots$ \\
\textsc{ra1h40dec-35} (``Mid-North'')				& [0, 50] 	& [-42, -28]	& 2021 Nov 30 -- 2022 Jan 02 	& 264 	& $\cdots$	& $\cdots$ \\
\textsc{ra1h40dec-35} (``Mid-North'')$^{\dagger}$	& [0, 50] 	& [-42, -28]	& 2022 Dec 01 -- 2022 Dec 24 	& 259 	& $\cdots$	& $\cdots$ \\
\textsc{ra12h30dec-35} (``Backside'')				& [150, 225]& [-42, -28]	& 2021 Feb 03 -- 2021 Mar 21 	& 339 	& (480) Hansa	& 3.5 \\
													&			&				&								&		& (1266) Tone	& 4.0 \\
\textsc{ra12h30dec-35} (``Backside'')				& [150, 225]& [-42, -28]	& 2022 Feb 14 -- 2022 Mar 21 	& 344 	& (36) Atalante	& 6.4 \\
													&			&				&								&		& (445) Edna	& 3.4 \\
													&			&				&								&		& (773) Irmintraud	& 8.3 \\
\textsc{ra12h30dec-35} (``Backside'')$^{\dagger}$	& [150, 225]& [-42, -28]	& 2023 Feb 14 -- 2023 Mar 21 	& 344 	& (426) Hippo	& 27.9 \\
													&			&				&								&		& (702) Alauda	& 11.0 \\
													&			&				&								&		& (705) Erminia	& 13.6 \\
													&			&				&								&		& (762) Pulcova	& 13.8 \\
\hline
\multicolumn{6}{l}{\footnotesize{$\dagger$ Planned future observations.}} \\
\multicolumn{6}{l}{\footnotesize{$\ddagger$ Only the northernmost 7.5$^\circ$ observed.}} \\
\label{tab:pred_3G}
\end{longtable}

\end{document}